\documentclass[nofootinbib,floatfix,onecolumn]{revtex4}

\usepackage{amsmath,ulem}
\usepackage{amssymb}
\usepackage{graphicx}
\usepackage{rotating}
\usepackage{color}
\usepackage{multirow} 
\usepackage[T1]{fontenc} 
\usepackage{slashed}
\usepackage[mathscr]{eucal}
\usepackage{color}
\usepackage{longtable}
\usepackage{ulem}

\newcommand{\MET}{$\slashed{E}_T$}
\newcommand{\EQ}{Eq.}
\newcommand{\EQs}{Eqs.}
\newcommand{\TABLE}{Table}
\newcommand{\TABLEs}{Tables}

\newcommand{\LLE}{$\hat{l}\hat{l}\hat{e}$}
\newcommand{\LQD}{$\hat{l}\hat{q}\hat{d}$}
\newcommand{\UDD}{$\hat{u}\hat{d}\hat{d}$}

\newcommand{\LQDspace}{$\hat{l}\hat{q}\hat{d} \;$}
\newcommand{\UDDspace}{$\hat{u}\hat{d}\hat{d} \;$}

\newcommand{\AddrWue}{%
Institut f\"ur Theoretische Physik und Astronomie, 
Universit\"at W\"urzburg\\
97074 W\"urzburg,Germany}

\newcommand{\AddrBonn}{%
Bethe Center for Theoretical Physics \& Physikalisches Institut der 
Universit\"at Bonn, \\
Nu{\ss}allee 12, 53115 Bonn, Germany
}

\newcommand{\AddrOrsay}{%
Laboratoire de Physique Th\'eorique, CNRS -- UMR 8627, Universit\'e de Paris-Sud 11\\ F-91405 Orsay Cedex, France}

\preprint{BONN-TH-2012-08,LPT-12-41}
\begin{document}

\title{General MSSM signatures at the LHC with and without $R$-parity}

\author{H. K. Dreiner}\email{dreiner@uni-bonn.de}
\author{F. Staub}\email{fnstaub@physik.uni-bonn.de}
\affiliation{\AddrBonn}

\author{A. Vicente} \email{avelino.vicente@th.u-psud.fr}
\affiliation{\AddrOrsay}

\author{W. Porod} \email{porod@physik.uni-wuerzburg.de}
\affiliation{\AddrWue}

\begin{abstract}
We present the possible signatures appearing in general realizations
of the MSSM based on 14 unrelated mass parameters at the SUSY
scale. The parameters of the general MSSM are reduced by assuming 
a degeneracy of the sfermions of the first two generations with the same 
quantum numbers. We also assume no mass-splitting between neutral and 
charged Higgsinos. We do allow for separate soft breaking terms for the 
third generation sfermons. We consider all possible resulting $14!\approx 9
\cdot10^{10}$ relevant mass orderings and check for the dominant decay 
cascades and the corresponding collider signatures. In determining
the dominant decay modes we assume that mixing between sparticles is 
sub-dominant. As preferred signatures, we consider charged leptons, 
missing transverse momentum, jets, and $W,\,Z$ or Higgs bosons. We 
include also the cases of bi- and trilinear $R$-parity violation and show 
that specific signatures can be used to distinguish the different scenarios.
\end{abstract}

\maketitle

\section{Introduction} 

The \texttt{ATLAS} \cite{Aad:2009wy} and \texttt{CMS}
\cite{Chatrchyan:2008zzk} experiments at the LHC have collected about 
5 fb${}^{-1}$ of data. Among other extensions of the Standard Model of
particle physics (SM) they have searched for supersymmetry (SUSY)
\cite{Nilles:1983ge,Martin:1997ns}, however so far to no avail. For 
the published 1\,fb$^{-1}$ data see
Refs.~\cite{Aad:2011ib,ATLAS:2011ad,Aad:2011cwa,Chatrchyan:2011ek,Chatrchyan:2011qs,Chatrchyan:2011zy}.
The more recent analyses have been presented in preliminary form in
Refs.~\cite{moriond}.  For a recent best-fit to the CMSSM, including
the most recent LHC data, see also \cite{Bechtle:2012zk,other}. The present
searches are largely based on the assumption of conserved $R$-parity
\cite{Farrar:1978xj}, where the lightest supersymmetric particle (LSP)
is stable. For cosmological reasons it must be electrically and color
neutral. All produced supersymmetric particles are typically expected
to cascade decay to the LSP promptly within the detector; the LSP
itself escapes detection leading to missing transverse energy,
$\not\!\!E_T$, as a typical signature. Thus most searches include a
more or less strict lower cut on $\not\!\!E_T$.

The minimal supersymmetric standard model (MSSM) has in its most
general form 124 free parameters \cite{Haber:1997if}. Most of these
arise from the supersymmetry breaking sector. In particular there is a
free mass parameter for each of the new supersymmetric particles. Thus
in the most general case any mass ordering of the spectrum is
possible. However it is difficult to get a meaningful interpretation
of the experimental searches in terms of 124 parameters. Thus
simplifying assumptions are made which dramatically reduce the number
of parameters. The most widely considered case is the CMSSM, often
also called the mSUGRA model. This has five free parameters. Other
popular reduced parameter scenarios include: gauge mediated SUSY
breaking (GMSB), anomaly
mediated SUSY breaking (AMSB) or mirage mediation. For the collider
phenomenology see for example
\cite{Giudice:1998bp,Paige:1999ui,Baer:2007eh,Conley:2011tq}.

The reduction to a significantly smaller set of free parameters comes
at the price of a significant loss of generality. The central point of
this paper is to analyse in general the possible mass orderings of the
spectrum and determine the resulting \textit{dominant} LHC signatures.
The dominant signature will be determined by the dominant decay modes
of the supersymmetric particles. We discuss these decays in detail
below and show our results for the dominant modes in
Tables~\ref{tab:signatures_colored}, \ref{tab:signatures_scalars} and
\ref{tab:signatures_fermion}. Consider the CMSSM as an example, which
has five free parameters. These are fixed at the unification scale
($\sim10^{16}\,$GeV). The low-energy supersymmetric spectrum and the
particlecouplings are computed using the renormalization group
equations (RGEs) via publically available codes
\cite{Paige:2003mg,Allanach:2001kg,Djouadi:2002ze,Porod:2003um}. This
leads to very specific features in the spectrum. For example typically
the gluinos and squarks are the heaviest particles and the
right-handed sleptons and the lightest neutralino are the lightest.
Overall 47 different mass hierarchies are possible
\cite{Konar:2010bi}.  This is a very small number compared to the
general MSSM. Within the general MSSM, if one assumes a degeneracy of
the sfermions with the same quantum numbers and no mass-splitting
between neutral and charged Higgsinos, there are nine mass
parameters. These lead to $9!  =362\,880$ mass hierarchies, as
discussed in Konar \textit{et al.}  \cite{Konar:2010bi}. We go one
step beyond this work and separate the soft-breaking terms of the
third generation sfermions from the other two generations resulting in
14 free parameters and $14!=87\,178\,291\,200\approx9\cdot10^{10}$
possible hierarchies.\footnote{See also related work in
  Ref. \cite{Berger:2008cq} where, in the same spirit of generality as
  in our work, complete scans of the pMSSM \cite{pMSSM} were
  performed.}

We extend the study \cite{Konar:2010bi} by several other aspects: we
allow for a mass-splitting between charged and neutral components of
Higgsinos and Winos. Since this breaks $SU(2)$ invariance, we
distinguish explicitly between charged leptons and missing transverse
energy ($\slashed{E}_T$) as a signature.  We start with the case of
$R$-parity conservation ($R$pC). Later we also consider bi- and
trilinear $R$-parity violating ($R$pV) parameters in detail, significantly 
going beyond the work in \cite{Konar:2010bi}. Our goal is to determine 
the widest possible set of distinct signatures, which we can still 
compute. These can then be employed to determine a broad 
based search for supersymmetry at the LHC.  They can also be used 
to find ways of distinguishing $R$pV from $R$pC.

The rest of the paper is organized as follows: In sec.~\ref{sec:model}
we give the basic definitions and conventions used throughout the
paper. We explain in detail our approach and the underlying
assumptions in sec.~\ref{sec:analysis}. In sec.~\ref{sec:results} we
discuss our results before we conclude in sec.~\ref{sec:conclusion}.

\section{Model definitions}
\label{sec:model}

In the following we consider the minimal supersymmetric standard model
(MSSM) \cite{Nilles:1983ge,Martin:1997ns}, extended to include
$R$-parity violation 
\cite{Hall:1983id,Dreiner:1997uz,Allanach:2003eb,Bhattacharyya:1997vv,Barger:1989rk,Allanach:1999ic,Hirsch:2000ef,Barbier:2004ez}. 
The MSSM particle content consists of the chiral superfields
$\hat{q}_a\, ({\bf 3},{\bf2 },\frac{1}{6})$, $\hat{\ell}_a\,({\bf
1},{\bf 2},-\frac{1}{2})$, $\hat {H}_d\, ({\bf 1},{\bf
2},-\frac{1}{2})$, $\hat{H}_u\,({\bf1},{\bf 2},
\frac{1}{2})$, $\hat{d}^c_a\,({\bf \bar{3}},{\bf1},\frac{1}{3})$, $
\hat{u}^c_a\,({\bf\bar{3}},{\bf 1},\frac{2}{3})$, $\hat{e}^c_a\, ({\bf
1},{\bf 1},1)$ and the vector superfields $\tilde{g}_\alpha \, ({\bf
8},{\bf 1},0)$, $\tilde{W}^i\, ({\bf 1},{\bf 3},0)$, $\tilde{B}\,({\bf
1},{\bf 1},0)$.  In parentheses we give the SM gauge quantum numbers
with respect to $SU(3)_c \times SU(2)_L \times U(1)_Y$. $a=1,2,3$ is a
generation index. The most general renormalizable and SM gauge invariant 
superpotential is
\begin{equation}
 W= Y^{a
   b}_{e}\,\hat{\ell}_a^j\,\hat{e}_b^c\,\hat{H}_d^i\,\epsilon_{ij}+Y^{a
   b}_{d}\,\hat{q}^{j \alpha}_a\,\hat{d}_{\alpha
   b}^c\,\hat{H}_d^i\,\epsilon_{ij} + Y^{a b}_{u} \hat{q}^{i
   \alpha}_a\,\hat{u}_{\alpha b}^c\,\hat{H}_u^j\,\epsilon_{ij} +
 \mu\,\hat{H}_u^i\,\hat{H}_d^j \epsilon_{ij} +
 W_{\slashed{R}} \thickspace .
\end{equation}
Here $a,b=1,2,3$ are generation indices and $i,j=1,2$ are $SU(2)_L$
gauge indices of the fundamental representation. $\epsilon_{ij}$ is
the totally anti-symmetric tensor. $Y_e,\,Y_d,\,Y_u$ are dimensionless
3x3 matrices of Yukawa couplings. $W_{\slashed{R}}$ contains the
well-known $R$-parity violating terms, which are discussed below. The
soft SUSY breaking potential reads
\begin{equation}
\mathscr{V}_{\text{SB}} =
\mathscr{V}_{\text{SB},R} + \mathscr{V}_{\text{SB},\slashed{R}}\,, 
\end{equation}
with the $R$pC part given in a hopefully clear notation 
based on the conventions of ref.~\cite{Allanach:2008qq}
\begin{eqnarray}
\nonumber \mathscr{V}_{\text{SB},R} &=& m_{H_u}^2 |H_u|^2 + m_{H_d}^2
|H_d|^2+ \tilde{q}^\dagger m_{\tilde{q}}^2 \tilde{q} +
\tilde{l}^\dagger m_{\tilde{l}}^2 \tilde{l} + \tilde{d}^\dagger
m_{\tilde{d}}^2 \tilde{d} + \tilde{u}^\dagger m_{\tilde{u}}^2
\tilde{u} \nonumber \\ && + \frac{1}{2}\left(M_1 \, \tilde{B}
\tilde{B} + M_2 \, \tilde{W}_i \tilde{W}^i + M_3 \, \tilde{g}_\alpha
\tilde{g}^\alpha + h.c.\right) \nonumber \\ && 
- H_u \tilde{q}
T_u\tilde{u}^\dagger +H_d \tilde{q} T_d \tilde{d}^\dagger + H_d
\tilde{l} T_e \tilde{e}^\dagger + B_\mu H_u H_d \thickspace .
\end{eqnarray}
 We do not specify the $R$-parity violating soft SUSY breaking soft
 terms in $\mathscr{V}_{\text{SB},\slashed{R}}$, for simplicity.

In general, the SUSY particles (sparticles) mix after electroweak
symmetry breaking (EWSB) giving rise to 28 mass eigenstates. (12
squarks, 6 charged sleptons, 3 neutral sleptons, 4 neutralinos, 2
charginos and 1 gluino; we identify all particles in $SU(3)_c\times
U(1)_{em}$ irreducible representations.)  With no \textit{a priori} model
explaining the masses, these 28 states lead to
$28!\simeq3\cdot10^{29}$ possible mass orderings or
\textit{hierarchies}. Unfortunately, it is computationally impossible
to classify this general setup. Therefore, we shall make some
assumptions, which we consider reasonable, to reduce the number of
hierarchies to a manageable amount. In the following we shall be
interested in \textit{dominant} effects on the spectrum and the
decays. We make two major simplifications:
\begin{enumerate}
\item[(i)] The mixing between sparticles is sub-dominant, so we can identify
the mass eigenstates with the corresponding gauge eigenstates.  The
only exception are the Higgsinos, which we assume to be maximally
mixed.
\item[(ii)] The first and second generations of sfermions of the same
kind are degenerate in mass. We consider the third generation masses as
independent parameters, \textit{e.g.} for the sleptons
\begin{eqnarray}
m_{\tilde e L}&=& m_{\tilde \mu L}=m_{\tilde\nu_e}=m_{\tilde\nu_\mu}=
m_{\tilde \ell,11} \\
m_{\tilde e R}&=& m_{\tilde \mu R}=m_{\tilde e,11} \\
m_{\tilde \tau L}&=& m_{\tilde\nu_\tau}=m_{\tilde \ell,33}\\
m_{\tilde\tau R}&=& m_{\tilde e,33}\,,
\end{eqnarray}
\end{enumerate}
and analogously for the squarks. These two assumptions leave us with
14 relevant mass parameters,
\begin{eqnarray}
& M_1, M_2, M_3, \mu & \\ & m_{\tilde{e},11}, m_{\tilde{e},33},
  m_{\tilde{\ell},11}, m_{\tilde{\ell},33} & \\ 
& m_{\tilde{d},11},
  m_{\tilde{d},33}, m_{\tilde{u},11}, m_{\tilde{u},33},
  m_{\tilde{q},11}, m_{\tilde{q},33} & \thickspace ,
\end{eqnarray}
and thus $14!$ different hierarchies.\footnote{In fact, as explained in the
next section, this will imply $2^4 \cdot 14!$ possibilities due to the
consideration of decay chains where we make the distinctions $\tilde{W
}^0/\tilde{W}^\pm$, $\tilde{H}^0/\tilde{H}^\pm$, $\tilde{\ell}/\tilde{
\nu}$ and $\tilde{\tau}/\tilde{\nu}_\tau$.} Furthermore, the identification of the first and
second generation sfermions allows us to reduce the number of fields
we need to take into account in our analysis. We combine them because,
by assumption, they lead to the same signatures:
\begin{eqnarray}
(\tilde{e}_L/\tilde{\mu}_L)&\rightarrow&\tilde{\ell}, \nonumber\\
(\tilde{e}_R/\tilde{\mu}_R) &\rightarrow& \tilde{e}, \nonumber\\ 
(\tilde{d}_L/\tilde{s}_L/\tilde{u}_L/\tilde{c}_L) &\rightarrow&\tilde{q}, \nonumber\\
(\tilde{d}_R/\tilde{s}_R) &\rightarrow& \tilde{d}, \nonumber\\
(\tilde{u}_R/\tilde{c}_R)&\rightarrow& \tilde{u}, \nonumber \\
(\tilde{\nu}_e/\tilde{\nu}_\mu) &\rightarrow& \tilde{\nu},
\end{eqnarray} 
as well as the two Higgsino-like neutralinos\footnote{We 
note that all signatures in our final results would appear $2^6 \cdot
4 = 256$ times more often if we consider all possible
combinations of the fields affected by this simplifying
assumption. However, since the relative importance will not change, we
do not include this factor to keep the numbers as low as possible.}. A
collection of the considered states as well as of the relevant mass
parameters is given in Table~\ref{tab:fields}.

\begin{table}[!bt]
\begin{tabular}{|l|c|c|}
\hline
Particle & Name & Mass \\
\hline
Bino-like neutralino & $\tilde{B}$  & $M_1$\\
Wino-like neutralino & $\tilde{W}^0$ & $M_2$ \\
Higgsino-like neutralinos & $\tilde{H}^0$ & $\mu$ \\
Gluino & $\tilde{G}$ & $M_3$\\
Wino-like chargino & $\tilde{W}^\pm$ & $M_2$ \\
Higgsino-like chargino & $\tilde{H}^\pm$  & $\mu$\\
left-Squarks (1./2. generation) & $\tilde{q}_{1,2} \equiv \tilde{q}$ & $m_{\tilde{q},11}$  \\  
down-right Squarks (1./2. generation) & $\tilde{d},\tilde{s} \equiv \tilde{d}$& $m_{\tilde{d},11}$ \\ 
up-right Squarks (1./2. generation) & $\tilde{u}, \tilde{c} \equiv \tilde{u}$  & $m_{\tilde{u},11}$\\  
left charged sleptons (1./2. generation) & $\tilde{e}_L, \tilde{\mu}_L \equiv \tilde{l}$ & $m_{\tilde{l},11}$ \\
sneutrinos (1./2. generation) & $\tilde{\nu}_e, \tilde{\nu}_\mu \equiv \tilde{\nu}$ & $m_{\tilde{l},11}$ \\  
right sleptons (1./2. generation) & $\tilde{e}_R, \tilde{\mu}_R \equiv \tilde{e}$  & $m_{\tilde{e},11}$\\
left-Squarks (3. generation) & $\tilde{q}_3$ & $m_{\tilde{q},33}$ \\  
down-right Squarks (3. generation) & $\tilde{b}$ & $m_{\tilde{d},33}$\\ 
up-right Squarks (3. generation) & $\tilde{t}$ & $m_{\tilde{u},33}$\\  
left staus (3. generation) & $\tilde{\tau}_L$ & $m_{\tilde{l},33}$\\
sneutrinos (3. generation) & $\tilde{\nu}_\tau$ & $m_{\tilde{l},33}$\\  
right sleptons (3. generation) & $\tilde{\tau}_R$ & $m_{\tilde{e},33}$\\
\hline 
\end{tabular}
\caption{Particle content and relevant mass parameters.}
\label{tab:fields}
\end{table}
Going beyond this, we also study the impact of the different $R$pV
superpotential couplings \cite{Hall:1983id,Allanach:2003eb,Dreiner:1997uz,Hirsch:2000ef,Barbier:2004ez}
\begin{equation} \label{rpv-superpot}
W_{\slashed{R}} = \epsilon_i \hat{\ell}_i \hat{H}_u +\frac{1}{2}  
\lambda_{ijk} \hat{\ell}_i \hat{\ell}_j \hat{e}_k^c  + \frac{1}{2}  
\lambda^{'}_{ijk} \hat{q}_i \hat{d}_j^c \hat{\ell}_k + \frac{1}{2} 
\lambda^{''}_{ijk} \hat{u}_i^c \hat{d}_j^c \hat{d}_k^c \,,
\end{equation}
where we have suppressed the $SU(2)_L$ and $SU(3)_c$ gauge indices.  The two
main phenomenological effects of $R$pV are
\begin{enumerate}
\item[(i)] SUSY particles can decay
directly to SM particles. In particular, the LSP is unstable. 
\item[(ii)] SUSY particles can be produced singly; resonant or in 
associated production \cite{Dimopoulos:1988fr,Hall:1983id,Dreiner:2000vf,Allanach:1997sa,Bernhardt:2008mz}.
\end{enumerate}
Furthermore, the additional interactions in Eq.~(\ref{rpv-superpot})
break either lepton (L) or baryon number (B). There is thus a large
set of experimental bounds on the parameters $\epsilon_i,\,\lambda,\,
\lambda',\,\lambda''$
\cite{Barger:1989rk,Dreiner:1997uz,Bhattacharyya:1997vv,Allanach:1999ic,Hirsch:2000ef,Dreiner:2001kc,Allanach:2003eb,Barbier:2004ez,Dreiner:2006gu,Kao:2009fg,Dreiner:2010ye,Dreiner:2012mx},
which are typically very strict. The Yukawa couplings are then
significantly smaller than the gauge couplings\footnote{Note that for
  large supersymmetric masses, the bounds on the $R$pV typically
  become weaker, although this is not always the case
  \cite{Dreiner:2012mx}. Bounds from neutrino masses on the
  L-violating couplings typically have a weaker mass dependence.}. For
our analysis of the mass hierarchies, the effect of the $R$pV
couplings in the RGEs is typically small, since the couplings are
small \cite{Allanach:2003eb,Allanach:1999ic,Allanach:1999mh}. However,
we consider here general mass hierarchies. Since the LSP is not
stable, any particle can now be the LSP.

The $R$pV couplings can also lead to additional decays of all SUSY
particles, see for example the decay tables in \cite{Allanach:2006st}.
However, since the couplings are small, we consider this effect
\textit{sub-dominant} for all sparticles, other than the LSP.

\section{Strategy for the analysis}
\label{sec:analysis}

In total, we have $14! = 87,178,291,200\approx 9\cdot10^{10}$ hierarchies. 
Each one can be denoted as a chain of fields in decreasing order of mass from 
left to right:
\begin{equation}
 i_1 \dots i_n C r_1 \dots r_m
\end{equation}
$C$ denotes the lightest colored particle (LCP), excluding the third
generation. So $C$ is the lightest of the four fields $\tilde{G},\,
\tilde {q},\,\tilde{d}$ and $\tilde{u}$. The particles $\{i_k\}$ ($i$
for irrelevant) are all heavier, and contain among others
the remaining colored particles, other than possible third generation
squarks. The particles $\{r_k\}$ ($r$ for relevant) are all lighter
than $C$ and are potentially involved in the cascade decay and thus
important for our analysis.  As in Ref.~\cite{Konar:2010bi}, we assume
that $C$ is the only directly produced particle at the LHC. We do not
impose any restrictions on the LSP, denoted $r_m$ above.

We are interested in the determination of the \textit{dominant decay chains} 
for all hierarchies. These are the decay chains $C \to r_i \to
\dots \to r_m = \text{LSP}$ that will dominantly happen at the LHC for
each hierarchy. In order to find them we apply the same algorithm as
in Ref.~\cite{Konar:2010bi}:
\begin{enumerate}
 \item Find those SUSY particles which are lighter than the LCP and
 have the largest coupling to it.\footnote{See
 \TABLEs~\ref{tab:signatures_colored}-\ref{tab:signatures_fermion} for
 the coupling strengths and the corresponding decay products.} In
 general, that may apply to more than one particle.
 \item For each of those particles, search for the
 lighter particles with the largest coupling to it. Again, several
 possibilities can exist and have to be considered independently.
 \item Iterate step 2 until the LSP is reached.
\end{enumerate}
In principle, one can have more than one dominant decay chain for a
given hierarchy. That situation would correspond to decay chains with
similar rates at the LHC. Once the dominant decay chains are found one
can determine their signature. These signatures, denoted here as
\textit{dominant signatures},\footnote{If different signatures can be
  the result of a given decay chain we have chosen the one with the
  largest number of charged leptons. See also the remarks at the end
  of this section. Therefore, both \textit{dominant signatures} and
  \textit{dominant and best visible signatures} will be used as
  equivalent concepts in the following.}  represent the experimentally
relevant result of our study. They are obtained by summing up the
decay products of all steps in the decay chain. These are given,
together with the coupling strengths, in
\TABLEs~\ref{tab:signatures_colored}-\ref{tab:signatures_fermion}. We
have considered as final state particles in our analysis
\begin{enumerate}
\item charged leptons ($l$), 
\item jets ($j$), 
\item massive
bosons ($v$) 
\item missing transverse energy ($\slashed{E}_T$) (neutrinos and 
neutralino and sneutrino LSP, for $R$pC).
\end{enumerate}
Note
that \textit{massive bosons} stands for both gauge and Higgs
bosons. 

\medskip

In Table \ref{tab:signatures_colored} we see for example that the
$SU(2)_L$ singlet down-like squark, $\tilde d$, can dominantly decay
to a gluino or a bino, if either is lighter. If not, it has a wide
range of decays which are all suppressed, in our sense, but must be
considered, if the dominant decays are kinematically
blocked. Similarly, the $\tilde e$ field only has an unsuppressed
decay to the bino, if allowed.  Whereas the $\tilde l$ has three
unsuppressed modes, to the bino, neutral and charged wino. The charged
Higgsino, $\tilde H^\pm$, has unsuppressed decays to the third
generation scalars due to the large Yukawa couplings.

\medskip

While the case of a neutral LSP might be favored because it could also
give a valid dark matter candidate, we have included explicitly also
the two possibilities of a charged or colored LSP since these are not
necessarily ruled out: the relic density of the SUSY LSP could be tiny
enough to be cosmologically negligible and dark matter is formed by
other fields like the axion
\cite{Preskill:1982cy,Abbott:1982af,Dine:1982ah} or the axino
\cite{Covi:1999ty}.  In case of the gravitino, which we have not
mentioned so far, it may be the LSP, and thus the lightest SUSY
particle discussed here would be the NLSP. In such scenarios, the NLSP
would decay outside the detector into a gravitino
\cite{Ellis:2003dn,Ellis:2006vu}. We will nevertheless call this
particle the LSP, since the real LSP would not be discovered at the LHC in
such a case.  We refer the interested reader to \cite{Steffen:2008qp}
for a summary of dark matter candidates in supersymmetry. For our
discussion in the following the nature of dark matter is not relevant.
However, we want to point out that detailed studies for a charged or a
colored LSP exist in the literature
\cite{Abazov:2008qu,Aaltonen:2009kea,Raby:1997bpa,Baer:1998pg,Raby:1998xr}
which have motivated us to discuss these scenarios here.

As for the dominant decay chains, one can have more than one dominant
signature for a given hierarchy. This would happen if two dominant
decay chains have different signatures. However, if two dominant decay
chains have the same signature, this signature is only counted once.

\begin{table}
\centering
\begin{tabular}{|l l c|l l c|l l c|} 
\hline 
transition & strength & signature & transition &  strength & signature  & transition &  strength & signature \\ 
\hline
$\tilde{G}\leftrightarrow\tilde{d}$ & not sup.& $j$&$\tilde{G}\leftrightarrow\tilde{q}$ & not sup.& $j$&$\tilde{G}\leftrightarrow\tilde{u}$ & not sup.& $j$\\ 
$\tilde{G}\leftrightarrow\tilde{W}^0$ & sup.& $2 j$&$\tilde{G}\leftrightarrow\tilde{W}^\pm$ & sup.& $2 j$&$\tilde{G}\leftrightarrow\tilde{e}$ & str. sup.& $2 j + l$\\ 
$\tilde{G}\leftrightarrow\tilde{H}^0$ & str. sup.& $2 j$&$\tilde{G}\leftrightarrow\tilde{H}^\pm$ & str. sup.& $2 j$&$\tilde{G}\leftrightarrow\tilde{l}$ & str. sup.& $2 j + l$\\ 
$\tilde{G}\leftrightarrow\tilde{\nu}$ & str. sup.& $2 j + l$&$\tilde{G}\leftrightarrow\tilde{t}$ & not sup.& $j$&$\tilde{G}\leftrightarrow\tilde{b}$ & not sup.& $j$\\ 
$\tilde{G}\leftrightarrow\tilde{q}_3$ & not sup.& $j$&$\tilde{G}\leftrightarrow\tilde{\tau}_R$ & str. sup.& $3 j$&$\tilde{G}\leftrightarrow\tilde{\tau}_L$ & str. sup.& $3 j$\\ 
$\tilde{G}\leftrightarrow\tilde{\nu}_\tau$ & str. sup.& $2 j + \slashed{E}_T$&$\tilde{G}\leftrightarrow\tilde{B}$ & sup.& $2 j$ & & &\\ 
\hline \hline
$\tilde{d}\leftrightarrow\tilde{G}$ & not sup.& $j$&$\tilde{d}\leftrightarrow\tilde{q}$ & sup.& $2 j$&$\tilde{d}\leftrightarrow\tilde{u}$ & sup.& $2 j$\\ 
$\tilde{d}\leftrightarrow\tilde{W}^0$ & sup.& $j$&$\tilde{d}\leftrightarrow\tilde{W}^\pm$ & sup.& $j$&$\tilde{d}\leftrightarrow\tilde{e}$ & sup.& $j + l$\\ 
$\tilde{d}\leftrightarrow\tilde{H}^0$ & sup.& $j$&$\tilde{d}\leftrightarrow\tilde{H}^\pm$ & sup.& $j$&$\tilde{d}\leftrightarrow\tilde{l}$ & sup.& $j + l$\\ 
$\tilde{d}\leftrightarrow\tilde{\nu}$ & sup.& $j + l$&$\tilde{d}\leftrightarrow\tilde{t}$ & sup.& $2 j$&$\tilde{d}\leftrightarrow\tilde{b}$ & sup.& $2 j$\\ 
$\tilde{d}\leftrightarrow\tilde{q}_3$ & sup.& $2 j$&$\tilde{d}\leftrightarrow\tilde{\tau}_R$ & sup.& $2 j$&$\tilde{d}\leftrightarrow\tilde{\tau}_L$ & sup.& $2 j$\\ 
$\tilde{d}\leftrightarrow\tilde{\nu}_\tau$ & sup.& $j + \slashed{E}_T$&$\tilde{d}\leftrightarrow\tilde{B}$ & not sup.& $j$ & & &\\ 
\hline 
$\tilde{q}\leftrightarrow\tilde{G}$ & not sup.& $j$&$\tilde{q}\leftrightarrow\tilde{d}$ & sup.& $2 j$&$\tilde{q}\leftrightarrow\tilde{u}$ & sup.& $2 j$\\ 
$\tilde{q}\leftrightarrow\tilde{W}^0$ & not sup.& $j$&$\tilde{q}\leftrightarrow\tilde{W}^\pm$ & not sup.& $j$&$\tilde{q}\leftrightarrow\tilde{e}$ & sup.& $j + l$\\ 
$\tilde{q}\leftrightarrow\tilde{H}^0$ & sup.& $j$&$\tilde{q}\leftrightarrow\tilde{H}^\pm$ & sup.& $j$&$\tilde{q}\leftrightarrow\tilde{l}$ & sup.& $j + l$\\ 
$\tilde{q}\leftrightarrow\tilde{\nu}$ & sup.& $j + l$&$\tilde{q}\leftrightarrow\tilde{t}$ & sup.& $2 j$&$\tilde{q}\leftrightarrow\tilde{b}$ & sup.& $2 j$\\ 
$\tilde{q}\leftrightarrow\tilde{q}_3$ & sup.& $2 j$&$\tilde{q}\leftrightarrow\tilde{\tau}_R$ & sup.& $2 j$&$\tilde{q}\leftrightarrow\tilde{\tau}_L$ & sup.& $j + \slashed{E}_T$\\ 
$\tilde{q}\leftrightarrow\tilde{\nu}_\tau$ & sup.& $j + \slashed{E}_T$&$\tilde{q}\leftrightarrow\tilde{B}$ & not sup.& $j$ & & &\\ 
\hline  \hline
$\tilde{u}\leftrightarrow\tilde{G}$ & not sup.& $j$&$\tilde{u}\leftrightarrow\tilde{d}$ & sup.& $2 j$&$\tilde{u}\leftrightarrow\tilde{q}$ & sup.& $2 j$\\ 
$\tilde{u}\leftrightarrow\tilde{W}^0$ & sup.& $j$&$\tilde{u}\leftrightarrow\tilde{W}^\pm$ & sup.& $j$&$\tilde{u}\leftrightarrow\tilde{e}$ & sup.& $j + l$\\ 
$\tilde{u}\leftrightarrow\tilde{H}^0$ & sup.& $j$&$\tilde{u}\leftrightarrow\tilde{H}^\pm$ & sup.& $j$&$\tilde{u}\leftrightarrow\tilde{l}$ & sup.& $j + l$\\ 
$\tilde{u}\leftrightarrow\tilde{\nu}$ & sup.& $j + l$&$\tilde{u}\leftrightarrow\tilde{t}$ & sup.& $2 j$&$\tilde{u}\leftrightarrow\tilde{b}$ & sup.& $2 j$\\ 
$\tilde{u}\leftrightarrow\tilde{q}_3$ & sup.& $2 j$&$\tilde{u}\leftrightarrow\tilde{\tau}_R$ & sup.& $2 j$&$\tilde{u}\leftrightarrow\tilde{\tau}_L$ & sup.& $2 j$\\ 
$\tilde{u}\leftrightarrow\tilde{\nu}_\tau$ & sup.& $j + \slashed{E}_T$&$\tilde{u}\leftrightarrow\tilde{B}$ & not sup.& $j$ & & &\\ 
\hline \hline
$\tilde{t}\leftrightarrow\tilde{G}$ & not sup.& $j$&$\tilde{t}\leftrightarrow\tilde{d}$ & sup.& $2 j$&$\tilde{t}\leftrightarrow\tilde{q}$ & sup.& $2 j$\\ 
$\tilde{t}\leftrightarrow\tilde{u}$ & sup.& $2 j$&$\tilde{t}\leftrightarrow\tilde{W}^0$ & sup.& $j$&$\tilde{t}\leftrightarrow\tilde{W}^\pm$ & sup.& $j$\\ 
$\tilde{t}\leftrightarrow\tilde{e}$ & sup.& $j + l$&$\tilde{t}\leftrightarrow\tilde{H}^0$ & not sup.& $j$&$\tilde{t}\leftrightarrow\tilde{H}^\pm$ & not sup.& $j$\\ 
$\tilde{t}\leftrightarrow\tilde{l}$ & sup.& $j + l$&$\tilde{t}\leftrightarrow\tilde{\nu}$ & sup.& $j + l$&$\tilde{t}\leftrightarrow\tilde{b}$ & sup.& $2 j$\\ 
$\tilde{t}\leftrightarrow\tilde{q}_3$ & sup.& $2 j$&$\tilde{t}\leftrightarrow\tilde{\tau}_R$ & sup.& $2 j$&$\tilde{t}\leftrightarrow\tilde{\tau}_L$ & sup.& $2 j$\\ 
$\tilde{t}\leftrightarrow\tilde{\nu}_\tau$ & sup.& $j + \slashed{E}_T$&$\tilde{t}\leftrightarrow\tilde{B}$ & not sup.& $j$ & & &\\ 
\hline \hline
$\tilde{b}\leftrightarrow\tilde{G}$ & not sup.& $j$&$\tilde{b}\leftrightarrow\tilde{d}$ & sup.& $2 j$&$\tilde{b}\leftrightarrow\tilde{q}$ & sup.& $2 j$\\ 
$\tilde{b}\leftrightarrow\tilde{u}$ & sup.& $2 j$&$\tilde{b}\leftrightarrow\tilde{W}^0$ & sup.& $j$&$\tilde{b}\leftrightarrow\tilde{W}^\pm$ & sup.& $j$\\ 
$\tilde{b}\leftrightarrow\tilde{e}$ & sup.& $j + l$&$\tilde{b}\leftrightarrow\tilde{H}^0$ & not sup.& $j$&$\tilde{b}\leftrightarrow\tilde{H}^\pm$ & not sup.& $j$\\ 
$\tilde{b}\leftrightarrow\tilde{l}$ & sup.& $j + l$&$\tilde{b}\leftrightarrow\tilde{\nu}$ & sup.& $j + l$&$\tilde{b}\leftrightarrow\tilde{t}$ & sup.& $2 j$\\ 
$\tilde{b}\leftrightarrow\tilde{q}_3$ & sup.& $2 j$&$\tilde{b}\leftrightarrow\tilde{\tau}_R$ & sup.& $2 j$&$\tilde{b}\leftrightarrow\tilde{\tau}_L$ & sup.& $2 j$\\ 
$\tilde{b}\leftrightarrow\tilde{\nu}_\tau$ & sup.& $j + \slashed{E}_T$&$\tilde{b}\leftrightarrow\tilde{B}$ & not sup.& $j$ & & &\\ 
\hline \hline
$\tilde{q}_3\leftrightarrow\tilde{G}$ & not sup.& $j$&$\tilde{q}_3\leftrightarrow\tilde{d}$ & sup.& $2 j$&$\tilde{q}_3\leftrightarrow\tilde{q}$ & sup.& $2 j$\\ 
$\tilde{q}_3\leftrightarrow\tilde{u}$ & sup.& $2 j$&$\tilde{q}_3\leftrightarrow\tilde{W}^0$ & not sup.& $j$&$\tilde{q}_3\leftrightarrow\tilde{W}^\pm$ & not sup.& $j$\\ 
$\tilde{q}_3\leftrightarrow\tilde{e}$ & sup.& $j + l$&$\tilde{q}_3\leftrightarrow\tilde{H}^0$ & not sup.& $j$&$\tilde{q}_3\leftrightarrow\tilde{H}^\pm$ & not sup.& $j$\\ 
$\tilde{q}_3\leftrightarrow\tilde{l}$ & sup.& $j + l$&$\tilde{q}_3\leftrightarrow\tilde{\nu}$ & sup.& $j + l$&$\tilde{q}_3\leftrightarrow\tilde{t}$ & sup.& $2 j$\\ 
$\tilde{q}_3\leftrightarrow\tilde{b}$ & sup.& $2 j$&$\tilde{q}_3\leftrightarrow\tilde{\tau}_R$ & sup.& $2 j$&$\tilde{q}_3\leftrightarrow\tilde{\tau}_L$ & sup.& $j + \slashed{E}_T$\\ 
$\tilde{q}_3\leftrightarrow\tilde{\nu}_\tau$ & sup.& $j + \slashed{E}_T$&$\tilde{q}_3\leftrightarrow\tilde{B}$ & not sup.& $j$ & & &\\ 
\hline
\end{tabular}
\caption{Interactions for colored particles. We have considered for 
our analysis charged lepton ($l$), jets ($j$), massive bosons ($v$)
and missing transversal energy ($\slashed{E}_T$) as signatures.}
\label{tab:signatures_colored}
\end{table}

\begin{table}
\centering
\begin{tabular}{|l l c|l l c|l l c|} 
\hline 
transition & strength & signature & transition &  strength & signature  & transition &  strength & signature \\ 
\hline
$\tilde{e}\leftrightarrow\tilde{G}$ & str. sup.& $2 j + l$&$\tilde{e}\leftrightarrow\tilde{d}$ & sup.& $j + l$&$\tilde{e}\leftrightarrow\tilde{q}$ & sup.& $j + l$\\ 
$\tilde{e}\leftrightarrow\tilde{u}$ & sup.& $j + l$&$\tilde{e}\leftrightarrow\tilde{W}^0$ & sup.& $l$&$\tilde{e}\leftrightarrow\tilde{W}^\pm$ & sup.& $\slashed{E}_T$\\ 
$\tilde{e}\leftrightarrow\tilde{H}^0$ & sup.& $l$&$\tilde{e}\leftrightarrow\tilde{H}^\pm$ & sup.& $\slashed{E}_T$&$\tilde{e}\leftrightarrow\tilde{l}$ & sup.& $2 l$\\ 
$\tilde{e}\leftrightarrow\tilde{\nu}$ & sup.& $l + \slashed{E}_T$&$\tilde{e}\leftrightarrow\tilde{t}$ & sup.& $j + l$&$\tilde{e}\leftrightarrow\tilde{b}$ & sup.& $j + l$\\ 
$\tilde{e}\leftrightarrow\tilde{q}_3$ & sup.& $j + l$&$\tilde{e}\leftrightarrow\tilde{\tau}_R$ & sup.& $j + l$&$\tilde{e}\leftrightarrow\tilde{\tau}_L$ & sup.& $j + l$\\ 
$\tilde{e}\leftrightarrow\tilde{\nu}_\tau$ & sup.& $l + \slashed{E}_T$&$\tilde{e}\leftrightarrow\tilde{B}$ & not sup.& $l$ & & &\\ 
\hline  \hline
$\tilde{l}\leftrightarrow\tilde{G}$ & str. sup.& $2 j + l$&$\tilde{l}\leftrightarrow\tilde{d}$ & sup.& $j + l$&$\tilde{l}\leftrightarrow\tilde{q}$ & sup.& $j + l$\\ 
$\tilde{l}\leftrightarrow\tilde{u}$ & sup.& $j + l$&$\tilde{l}\leftrightarrow\tilde{W}^0$ & not sup.& $l$&$\tilde{l}\leftrightarrow\tilde{W}^\pm$ & not sup.& $\slashed{E}_T$\\ 
$\tilde{l}\leftrightarrow\tilde{e}$ & sup.& $2 l$&$\tilde{l}\leftrightarrow\tilde{H}^0$ & sup.& $l$&$\tilde{l}\leftrightarrow\tilde{H}^\pm$ & str. sup.& $\slashed{E}_T$\\ 
$\tilde{l}\leftrightarrow\tilde{t}$ & sup.& $j + l$&$\tilde{l}\leftrightarrow\tilde{b}$ & sup.& $j + l$&$\tilde{l}\leftrightarrow\tilde{q}_3$ & sup.& $j + l$\\ 
$\tilde{l}\leftrightarrow\tilde{\tau}_R$ & sup.& $j + l$&$\tilde{l}\leftrightarrow\tilde{\tau}_L$ & sup.& $j + l$&$\tilde{l}\leftrightarrow\tilde{\nu}_\tau$ & sup.& $l + \slashed{E}_T$\\ 
$\tilde{l}\leftrightarrow\tilde{B}$ & not sup.& $l$ & & & & & &\\ 
\hline \hline
$\tilde{\nu}\leftrightarrow\tilde{G}$ & str. sup.& $2 j + l$&$\tilde{\nu}\leftrightarrow\tilde{d}$ & sup.& $j + l$&$\tilde{\nu}\leftrightarrow\tilde{q}$ & sup.& $j + l$\\ 
$\tilde{\nu}\leftrightarrow\tilde{u}$ & sup.& $j + l$&$\tilde{\nu}\leftrightarrow\tilde{W}^0$ & not sup.& $\slashed{E}_T$&$\tilde{\nu}\leftrightarrow\tilde{W}^\pm$ & not sup.& $l$\\ 
$\tilde{\nu}\leftrightarrow\tilde{e}$ & sup.& $l + \slashed{E}_T$&$\tilde{\nu}\leftrightarrow\tilde{H}^0$ & str. sup.& $\slashed{E}_T$&$\tilde{\nu}\leftrightarrow\tilde{H}^\pm$ & sup.& $l$\\ 
$\tilde{\nu}\leftrightarrow\tilde{t}$ & sup.& $j + l$&$\tilde{\nu}\leftrightarrow\tilde{b}$ & sup.& $j + l$&$\tilde{\nu}\leftrightarrow\tilde{q}_3$ & sup.& $j + l$\\ 
$\tilde{\nu}\leftrightarrow\tilde{\tau}_R$ & sup.& $j + \slashed{E}_T$&$\tilde{\nu}\leftrightarrow\tilde{\tau}_L$ & sup.& $j + \slashed{E}_T$&$\tilde{\nu}\leftrightarrow\tilde{\nu}_\tau$ & sup.& $j + l$\\ 
$\tilde{\nu}\leftrightarrow\tilde{B}$ & not sup.& $\slashed{E}_T$ & & & & & &\\ 
\hline \hline
$\tilde{\tau}_R\leftrightarrow\tilde{G}$ & str. sup.& $3 j$&$\tilde{\tau}_R\leftrightarrow\tilde{d}$ & sup.& $2 j$&$\tilde{\tau}_R\leftrightarrow\tilde{q}$ & sup.& $2 j$\\ 
$\tilde{\tau}_R\leftrightarrow\tilde{u}$ & sup.& $2 j$&$\tilde{\tau}_R\leftrightarrow\tilde{W}^0$ & sup.& $j$&$\tilde{\tau}_R\leftrightarrow\tilde{W}^\pm$ & sup.& $\slashed{E}_T$\\ 
$\tilde{\tau}_R\leftrightarrow\tilde{e}$ & sup.& $j + l$&$\tilde{\tau}_R\leftrightarrow\tilde{H}^0$ & not sup.& $j$&$\tilde{\tau}_R\leftrightarrow\tilde{H}^\pm$ & not sup.& $\slashed{E}_T$\\ 
$\tilde{\tau}_R\leftrightarrow\tilde{l}$ & sup.& $j + l$&$\tilde{\tau}_R\leftrightarrow\tilde{\nu}$ & sup.& $j + \slashed{E}_T$&$\tilde{\tau}_R\leftrightarrow\tilde{t}$ & sup.& $2 j$\\ 
$\tilde{\tau}_R\leftrightarrow\tilde{b}$ & sup.& $2 j$&$\tilde{\tau}_R\leftrightarrow\tilde{q}_3$ & sup.& $2 j$&$\tilde{\tau}_R\leftrightarrow\tilde{\tau}_L$ & sup.& $2 j$\\ 
$\tilde{\tau}_R\leftrightarrow\tilde{\nu}_\tau$ & sup.& $j + \slashed{E}_T$&$\tilde{\tau}_R\leftrightarrow\tilde{B}$ & not sup.& $j$ & & &\\ 
\hline  \hline
$\tilde{\tau}_L\leftrightarrow\tilde{G}$ & str. sup.& $3 j$&$\tilde{\tau}_L\leftrightarrow\tilde{d}$ & sup.& $2 j$&$\tilde{\tau}_L\leftrightarrow\tilde{q}$ & sup.& $j + \slashed{E}_T$\\ 
$\tilde{\tau}_L\leftrightarrow\tilde{u}$ & sup.& $2 j$&$\tilde{\tau}_L\leftrightarrow\tilde{W}^0$ & not sup.& $j$&$\tilde{\tau}_L\leftrightarrow\tilde{W}^\pm$ & not sup.& $\slashed{E}_T$\\ 
$\tilde{\tau}_L\leftrightarrow\tilde{e}$ & sup.& $j + l$&$\tilde{\tau}_L\leftrightarrow\tilde{H}^0$ & not sup.& $j$&$\tilde{\tau}_L\leftrightarrow\tilde{H}^\pm$ & sup.& $\slashed{E}_T$\\ 
$\tilde{\tau}_L\leftrightarrow\tilde{l}$ & sup.& $j + l$&$\tilde{\tau}_L\leftrightarrow\tilde{\nu}$ & sup.& $j + \slashed{E}_T$&$\tilde{\tau}_L\leftrightarrow\tilde{t}$ & sup.& $2 j$\\ 
$\tilde{\tau}_L\leftrightarrow\tilde{b}$ & sup.& $2 j$&$\tilde{\tau}_L\leftrightarrow\tilde{q}_3$ & sup.& $j + \slashed{E}_T$&$\tilde{\tau}_L\leftrightarrow\tilde{\tau}_R$ & sup.& $2 j$\\ 
$\tilde{\tau}_L\leftrightarrow\tilde{B}$ & not sup.& $j$ & &  & & & &\\ 
\hline \hline
$\tilde{\nu}_\tau\leftrightarrow\tilde{G}$ & str. sup.& $2 j + \slashed{E}_T$&$\tilde{\nu}_\tau\leftrightarrow\tilde{d}$ & sup.& $j + \slashed{E}_T$&$\tilde{\nu}_\tau\leftrightarrow\tilde{q}$ & sup.& $j + \slashed{E}_T$\\ 
$\tilde{\nu}_\tau\leftrightarrow\tilde{u}$ & sup.& $j + \slashed{E}_T$&$\tilde{\nu}_\tau\leftrightarrow\tilde{W}^0$ & not sup.& $\slashed{E}_T$&$\tilde{\nu}_\tau\leftrightarrow\tilde{W}^\pm$ & not sup.& $j$\\ 
$\tilde{\nu}_\tau\leftrightarrow\tilde{e}$ & sup.& $l + \slashed{E}_T$&$\tilde{\nu}_\tau\leftrightarrow\tilde{H}^0$ & sup.& $\slashed{E}_T$&$\tilde{\nu}_\tau\leftrightarrow\tilde{H}^\pm$ & not sup.& $j$\\ 
$\tilde{\nu}_\tau\leftrightarrow\tilde{l}$ & sup.& $l + \slashed{E}_T$&$\tilde{\nu}_\tau\leftrightarrow\tilde{\nu}$ & sup.& $j + l$&$\tilde{\nu}_\tau\leftrightarrow\tilde{t}$ & sup.& $j + \slashed{E}_T$\\ 
$\tilde{\nu}_\tau\leftrightarrow\tilde{b}$ & sup.& $j + \slashed{E}_T$&$\tilde{\nu}_\tau\leftrightarrow\tilde{q}_3$ & sup.& $j + \slashed{E}_T$&$\tilde{\nu}_\tau\leftrightarrow\tilde{\tau}_R$ & sup.& $j + \slashed{E}_T$\\ 
$\tilde{\nu}_\tau\leftrightarrow\tilde{B}$ & not sup.& $\slashed{E}_T$ & & & & & & \\ 
\hline \hline
\end{tabular}
\caption{Interactions for uncolored scalars. The same definitions as in 
Table~\ref{tab:signatures_colored} are used. }
\label{tab:signatures_scalars}
\end{table}

\begin{table}
\centering
\begin{tabular}{|l l c|l l c|l l c|} 
\hline 
transition & strength & signature & transition &  strength & signature  & transition &  strength & signature \\ 
\hline
$\tilde{W}^0\leftrightarrow\tilde{G}$ & sup.& $2 j$&$\tilde{W}^0\leftrightarrow\tilde{d}$ & sup.& $j$&$\tilde{W}^0\leftrightarrow\tilde{q}$ & not sup.& $j$\\ 
$\tilde{W}^0\leftrightarrow\tilde{u}$ & sup.& $j$&$\tilde{W}^0\leftrightarrow\tilde{e}$ & sup.& $l$&$\tilde{W}^0\leftrightarrow\tilde{H}^0$ & not sup.& $v$\\ 
$\tilde{W}^0\leftrightarrow\tilde{H}^\pm$ & not sup.& $v$&$\tilde{W}^0\leftrightarrow\tilde{l}$ & not sup.& $l$&$\tilde{W}^0\leftrightarrow\tilde{\nu}$ & not sup.& $\slashed{E}_T$\\ 
$\tilde{W}^0\leftrightarrow\tilde{t}$ & sup.& $j$&$\tilde{W}^0\leftrightarrow\tilde{b}$ & sup.& $j$&$\tilde{W}^0\leftrightarrow\tilde{q}_3$ & not sup.& $j$\\ 
$\tilde{W}^0\leftrightarrow\tilde{\tau}_R$ & sup.& $j$&$\tilde{W}^0\leftrightarrow\tilde{\tau}_L$ & not sup.& $j$&$\tilde{W}^0\leftrightarrow\tilde{\nu}_\tau$ & not sup.& $\slashed{E}_T$\\ 
$\tilde{W}^0\leftrightarrow\tilde{B}$ & sup.& $2 l$ & &  & & & &\\ 
\hline \hline
$\tilde{W}^\pm\leftrightarrow\tilde{G}$ & sup.& $2 j$&$\tilde{W}^\pm\leftrightarrow\tilde{d}$ & sup.& $j$&$\tilde{W}^\pm\leftrightarrow\tilde{q}$ & not sup.& $j$\\ 
$\tilde{W}^\pm\leftrightarrow\tilde{u}$ & sup.& $j$&$\tilde{W}^\pm\leftrightarrow\tilde{e}$ & sup.& $\slashed{E}_T$&$\tilde{W}^\pm\leftrightarrow\tilde{H}^0$ & not sup.& $v$\\ 
$\tilde{W}^\pm\leftrightarrow\tilde{H}^\pm$ & not sup.& $v$&$\tilde{W}^\pm\leftrightarrow\tilde{l}$ & not sup.& $\slashed{E}_T$&$\tilde{W}^\pm\leftrightarrow\tilde{\nu}$ & not sup.& $l$\\ 
$\tilde{W}^\pm\leftrightarrow\tilde{t}$ & sup.& $j$&$\tilde{W}^\pm\leftrightarrow\tilde{b}$ & sup.& $j$&$\tilde{W}^\pm\leftrightarrow\tilde{q}_3$ & not sup.& $j$\\ 
$\tilde{W}^\pm\leftrightarrow\tilde{\tau}_R$ & sup.& $\slashed{E}_T$&$\tilde{W}^\pm\leftrightarrow\tilde{\tau}_L$ & not sup.& $\slashed{E}_T$&$\tilde{W}^\pm\leftrightarrow\tilde{\nu}_\tau$ & not sup.& $j$\\ 
$\tilde{W}^\pm\leftrightarrow\tilde{B}$ & sup.& $l + \slashed{E}_T$ & & & &  & &\\ 
\hline \hline
$\tilde{H}^0\leftrightarrow\tilde{G}$ & str. sup.& $2 j$&$\tilde{H}^0\leftrightarrow\tilde{d}$ & sup.& $j$&$\tilde{H}^0\leftrightarrow\tilde{q}$ & sup.& $j$\\ 
$\tilde{H}^0\leftrightarrow\tilde{u}$ & sup.& $j$&$\tilde{H}^0\leftrightarrow\tilde{W}^0$ & not sup.& $v$&$\tilde{H}^0\leftrightarrow\tilde{W}^\pm$ & not sup.& $v$\\ 
$\tilde{H}^0\leftrightarrow\tilde{e}$ & sup.& $l$&$\tilde{H}^0\leftrightarrow\tilde{l}$ & sup.& $l$&$\tilde{H}^0\leftrightarrow\tilde{\nu}$ & str. sup.& $\slashed{E}_T$\\ 
$\tilde{H}^0\leftrightarrow\tilde{t}$ & not sup.& $j$&$\tilde{H}^0\leftrightarrow\tilde{b}$ & not sup.& $j$&$\tilde{H}^0\leftrightarrow\tilde{q}_3$ & not sup.& $j$\\ 
$\tilde{H}^0\leftrightarrow\tilde{\tau}_R$ & not sup.& $j$&$\tilde{H}^0\leftrightarrow\tilde{\tau}_L$ & not sup.& $j$&$\tilde{H}^0\leftrightarrow\tilde{\nu}_\tau$ & sup.& $\slashed{E}_T$\\ 
$\tilde{H}^0\leftrightarrow\tilde{B}$ & not sup.& $v$ & & & & & &\\ 
\hline \hline
$\tilde{H}^\pm\leftrightarrow\tilde{G}$ & str. sup.& $2 j$&$\tilde{H}^\pm\leftrightarrow\tilde{d}$ & sup.& $j$&$\tilde{H}^\pm\leftrightarrow\tilde{q}$ & sup.& $j$\\ 
$\tilde{H}^\pm\leftrightarrow\tilde{u}$ & sup.& $j$&$\tilde{H}^\pm\leftrightarrow\tilde{W}^0$ & not sup.& $v$&$\tilde{H}^\pm\leftrightarrow\tilde{W}^\pm$ & not sup.& $v$\\ 
$\tilde{H}^\pm\leftrightarrow\tilde{e}$ & sup.& $\slashed{E}_T$&$\tilde{H}^\pm\leftrightarrow\tilde{l}$ & str. sup.& $\slashed{E}_T$&$\tilde{H}^\pm\leftrightarrow\tilde{\nu}$ & sup.& $l$\\ 
$\tilde{H}^\pm\leftrightarrow\tilde{t}$ & not sup.& $j$&$\tilde{H}^\pm\leftrightarrow\tilde{b}$ & not sup.& $j$&$\tilde{H}^\pm\leftrightarrow\tilde{q}_3$ & not sup.& $j$\\ 
$\tilde{H}^\pm\leftrightarrow\tilde{\tau}_R$ & not sup.& $\slashed{E}_T$&$\tilde{H}^\pm\leftrightarrow\tilde{\tau}_L$ & sup.& $\slashed{E}_T$&$\tilde{H}^\pm\leftrightarrow\tilde{\nu}_\tau$ & not sup.& $j$\\ 
$\tilde{H}^\pm\leftrightarrow\tilde{B}$ & not sup.& $v$ & & & & & &\\ 
\hline \hline
$\tilde{B}\leftrightarrow\tilde{G}$ & sup.& $2 j$&$\tilde{B}\leftrightarrow\tilde{d}$ & not sup.& $j$&$\tilde{B}\leftrightarrow\tilde{q}$ & not sup.& $j$\\ 
$\tilde{B}\leftrightarrow\tilde{u}$ & not sup.& $j$&$\tilde{B}\leftrightarrow\tilde{W}^0$ & sup.& $2 l$&$\tilde{B}\leftrightarrow\tilde{W}^\pm$ & sup.& $l + \slashed{E}_T$\\ 
$\tilde{B}\leftrightarrow\tilde{e}$ & not sup.& $l$&$\tilde{B}\leftrightarrow\tilde{H}^0$ & not sup.& $v$&$\tilde{B}\leftrightarrow\tilde{H}^\pm$ & not sup.& $v$\\ 
$\tilde{B}\leftrightarrow\tilde{l}$ & not sup.& $l$&$\tilde{B}\leftrightarrow\tilde{\nu}$ & not sup.& $\slashed{E}_T$&$\tilde{B}\leftrightarrow\tilde{t}$ & not sup.& $j$\\ 
$\tilde{B}\leftrightarrow\tilde{b}$ & not sup.& $j$&$\tilde{B}\leftrightarrow\tilde{q}_3$ & not sup.& $j$&$\tilde{B}\leftrightarrow\tilde{\tau}_R$ & not sup.& $j$\\ 
$\tilde{B}\leftrightarrow\tilde{\tau}_L$ & not sup.& $j$&$\tilde{B}\leftrightarrow\tilde{\nu}_\tau$ & not sup.& $\slashed{E}_T$ & & &\\ 
\hline 
\end{tabular}
\caption{Interactions for uncolored fermions. The same definitions as in 
Table~\ref{tab:signatures_colored} are used. }
\label{tab:signatures_fermion}
\end{table}

We exemplify this method with the hierarchy
\begin{equation}
i_1 \dots i_8 \tilde{G} \tilde{b} \tilde{H}^0 \tilde{W}^0 \tilde{l} \tilde{B}
\end{equation}
For the first two transitions only one possibility exists because the
largest couplings are $\tilde{G}\rightarrow\tilde{b}$ and $\tilde{b}
\rightarrow\tilde{H}^ 0$. However, the higgsino couples with the same
strength to the wino and to the bino, and therefore both branches have
to be considered.  Moreover, the wino will always take the way via the
slepton to decay into the LSP. In conclusion, there are two dominant
decay chains and two different dominant signatures:
\begin{align}
 \tilde{G} \rightarrow \tilde{b} \rightarrow \tilde{H}^0 \rightarrow 
\tilde{B} &: \hspace{0.3cm} 2 j + v\\
 \tilde{G} \rightarrow \tilde{b} \rightarrow \tilde{H}^0 \rightarrow 
\tilde{W}^0 \rightarrow \tilde{l} \rightarrow \tilde{B} &: \hspace{0.3cm}
  2 j + v+ 2 l
\end{align}
Both dominant signatures will appear $8!$ times because of all possible permutations of the fields heavier than the gluino LCP.

This method of classifying all possible signatures will be applied to
two different cases: (i) $R$-parity conservation and (ii) $R$-parity
violation. In the first case we will assume that all couplings in the
superpotential \EQ~\eqref{rpv-superpot} vanish exactly and thus the
LSP is stable. If neutral, it will contribute to the signature as
missing transverse energy, $\slashed{E}_T$. If colored or
electrically charged, the LSP interacts with the detector, leading to
visible tracks. Thus it cannot be regarded as $\slashed{E}_T$. We 
do not consider a stable massive electrically or colored charged particle 
as a separate signature.

In the second case, $R$-parity violation, the LSP is no longer
stable. Here we shall assume it decays in the detector. However, due
to the typically small $R$pV couplings, that is the only step of the
decay chain where these coupling play a role.  The rest is exactly the
same as in the $R$-parity conserving case.  Furthermore, we will
assume that one $R$pV coupling dominates the decay of the
LSP. Therefore, we treat separately four $R$-parity violating
scenarios: $\epsilon,\,\lambda,\,\lambda'$ or $\lambda''$
dominance.\footnote{We assume the LSP couples to the dominant $R$pV
operator, see also
\cite{Allanach:2003eb,Allanach:2006st}.} Our choices for the $R$-parity
violating decay modes are given in \TABLE~\ref{tab:RpVmodes}.
For completeness we note that in case of $R$pV the couplings can be so
small that displaced vertices can be measured at the LHC
\cite{deCampos:2007bn,deCampos:2008av,DeCampos:2010yu,Desch:2010gi,Dreiner:2011ft,kaplan2012}.
 However, as this depends on
the details of the model we did not include it as a signature here.
\begin{table}
  \begin{tabular}{|c|cccc|}
   \hline 
                   & $\epsilon$ & $\lambda$ & $\lambda'$ & $\lambda''$ \\
 \hline
 $\tilde{B}$ & $h^0 \nu$ & $l^+ l^- \nu$ & $l^\pm q \bar{q}'$ & $q q' q''$ \\
 $\tilde{W}^\pm$ & $Z^0 l^\pm$ & $3 l^\pm$ & $l^\pm q \bar{q}$ & $q q' q''$ \\
 $\tilde{W}^0$ & $W^\pm l^\mp$ & $l^+ l^- \nu$ & $l^\pm q \bar{q}'$ & $q q' q''$ \\
 $\tilde{G}$ & $q \bar{q}' l^\pm$ & $q \bar{q} l^+ l^- \nu$ & $l^\pm q \bar{q}'$ & $q q' q''$ \\
 $\tilde{H}^\pm$ & $Z^0 l^\pm$ & $3 l^\pm$ & $l^\pm q \bar{q}$ & $q q' q''$ \\
 $\tilde{H}^0$ & $W^\pm l^\mp$ &$l^+ l^- \nu$ & $l^\pm q \bar{q}'$ & $q q' q''$ \\
 $\tilde{q}$ & $l^\pm q$ & $q l^+ l^- \nu$ & $l^\pm q$ & $4 q$ \\
 $\tilde{d}$ & $l^\pm q$ & $q l^+ l^- \nu$ & $l^\pm q$ & $q q'$ \\
 $\tilde{u}$ & $q \nu$ & $q l^+ l^- \nu$ & $l^\pm q \bar{q}' q''$ & $q q'$ \\
$\tilde{l}$ & $q \bar{q}'$ & $l^\pm \nu$ & $q \bar{q}'$ & $q q' q'' l^\pm$ \\
$\tilde{\nu}$ & $q \bar{q}$ & $l^+ l^-$ & $q \bar{q}$ & $q q' q'' \nu$ \\
$\tilde{e}$ & $l^\pm \nu$ & $l^\pm \nu$ & $l^\pm l^\pm q \bar{q}'$ & $q q' q'' l^\pm$ \\
$\tilde{q}_3$ & $l^\pm q$ & $q l^+ l^- \nu$ & $l^\pm q$ & $4 q$ \\
$\tilde{b}_R$ & $q \nu$ & $q l^+ l^- \nu$ & $q \nu$ & $q q'$ \\
$\tilde{t}_R$ & $l^\pm q$ & $q l^+ l^- \nu$ & $l^\pm q \bar{q}' q''$ & $q q'$ \\
$\tilde{\tau}_L$ & $q \bar{q}'$ & $l^\pm \nu$ & $q \bar{q}'$ & $q q' q'' \tau$ \\
$\tilde{\nu}_\tau$  & $q \bar{q}$ & $l^+ l^-$ & $q \bar{q}$ & $q q' q'' \nu$ \\
$\tilde{\tau}_R$ & $\tau \nu$ & $l^\pm \nu$ & $l^\pm \nu q \bar{q}$ & $q q' q'' \tau$ \\
\hline
\end{tabular}
\caption{$R$-parity violating decay modes of the LSP
\cite{Dreiner:1997uz,Porod:2000hv,Hirsch:2003fe,Barbier:2004ez}. Note that
we have chosen charged lepton final states over \MET\ and thus neglected the
decay $\tilde B\rightarrow\nu q\bar q'$, for example.}
\label{tab:RpVmodes}
\end{table}

Some subtleties may arise when finding the dominant decay chains. Our
choice for the coupling strengths and the decay products for the
different steps of the decay chains also contain some additional
assumptions. Therefore, some remarks are required:

\begin{itemize}

\item We distinguish $\tilde{W}^0/\tilde{W}^\pm$,$\,\tilde{H}^0/\tilde
  {H}^\pm$, $\tilde{l}/\tilde{\nu}$ and $\tilde{\tau}/\tilde{\nu}_
  \tau$ in the decay chains. This is beyond what has been done in
  \cite{Konar:2010bi}, and is motivated by the fact that we
  distinguish between charged leptons and $\slashed{E}_T$ as a
  signature. Moreover, the charge of the LSP is correlated with the
  charge of the particles produced in the decay chain. Hence, it is
  not possible to choose them arbitrarily as charged lepton or
  $\slashed{E}_T$ when restricting to a chargeless LSP. The easiest
  example to illustrate this is the $\tilde{B} \to \tilde{L}$
  transition, where {$\tilde{L}$} is a scalar $SU(2)_L$ doublet
  containing a charged slepton and a sneutrino. The decay product of
  the transition can be either a charged lepton, for $\tilde{L} \equiv
  \tilde l$, or $\slashed{E}_T$, for $\tilde{L}
  \equiv\tilde\nu$. Similar correlations are important in order to
  correctly include the $R$pV decays. For instance, in the
  $\lambda$-dominated scenario a $\tilde{W}^\pm$ LSP will decay into
  $3 l$. (We consider the decay to 2$\nu+l$ less visible.)  On the
  other hand, a $\tilde{W}^0$ LSP will decay into $\slashed{E}_T + 2
  l$. This distinction between fields of different electric charge
  leads to $16 \cdot 14!= 1,394,852,659,200\approx10^{12}$ different field
  insertions in the $14!$ hierarchies. We also note that this
  distinction is not required for squarks, since their electric charge
  is reabsorbed in that of the jets.

\item Emitted $\tau$'s are regarded as ordinary jets.

\item When, for a given transition, two different decay products with
  similar strengths are possible, we always choose the one with the
  largest amount of charged leptons. When the choice is between
  $\tau$s and $\slashed{E}_T$, we have always chosen
  $\slashed{E}_T$. For example, the transition $\tilde{B}
\leftrightarrow \tilde{W}^0$ can either emit two jets, two charged
leptons or two neutrinos, and we choose two charged leptons.

\item We disregard the possibility of degeneracies among fields of
  different types (with the exceptions mentioned above concerning
  first and second generation sfermions). Therefore, 2-body decays
  have no phase space suppression.

\item We do not treat jets originating from third generation quarks
  separately. 

\end{itemize}

\section{Results}
\label{sec:results}
\subsection{Introduction and $R$pC Case}
\begin{table}[hbt]
\centering
\begin{tabular}{| c | c | c | c | c | c | c | c | c | c |} 
\hline 
& \multicolumn{3}{c|}{$n_v=0$} & \multicolumn{3}{c|}{$n_v=1$} & \multicolumn{3}{c|}{$n_v=2$} \\ \hline 
$n_l$ & $n_j=1$ & $n_j=2$ & $n_j>2$ & $n_j=1$ & $n_j=2$ & $n_j>2$& $n_j=1$ & $n_j=2$ & $n_j>2$ \\ \hline 
0 & $31.16$ & $11.76$ & $16.48$ & $4.03$ & $1.82$ & $1.27$ & $0.39$ & $0.11$ & $0.02$\\ 
\hline 
1 & $6.69$ & $3.47$ & $5.72$ & $0.42$ & $0.1$ & $0.36$ & $0.02$ & $7\cdot 10^{-3}$ & $0.02$\\ 
\hline 
2 & $6.5$ & $1.85$ & $5.09$ & $0.22$ & $0.11$ & $0.23$ & $5\cdot 10^{-3}$ & $2\cdot 10^{-3}$ & $7\cdot 10^{-3}$\\ 
\hline 
3 & $0.16$ & $0.25$ & $1.09$ & $0.02$ & $4\cdot 10^{-3}$ & $0.03$ & $2\cdot 10^{-4}$ & $2\cdot 10^{-4}$ & $2\cdot 10^{-3}$\\ 
\hline 
4 & $0.25$ & $0.05$ & $0.29$ & $4\cdot 10^{-3}$ & $2\cdot 10^{-3}$ & $8\cdot 10^{-3}$ & $3\cdot 10^{-4}$ & $5\cdot 10^{-5}$ & $5\cdot 10^{-4}$\\ 
\hline 
\end{tabular} 
\caption{Results for $R$-parity conservation and a neutral LSP.  $n_v$
  denotes the number of bosons, $n_j$ the number of jets and $n_l$ the
  number of charged leptons from the single cascade chain. All numbers
  in this table refer to percentages of a specific signature.}
\label{tab:res-rpc-neutral}
\end{table}
We present our results in terms of how large the percentage of models
is leading to the signature considered.  We have categorized the
signatures by the nature of the LSP: (a) neutral ($\tilde{B}$,
$\tilde{W}^0$, $\tilde{H}^0$, $\tilde{\nu}$, $\tilde{\nu}_\tau$), (b)
charged ($\tilde{l}$, $\tilde{e}$, $\tilde{ \tau}_L$,
$\tilde{\tau}_R$, $\tilde{W}^+$, $\tilde{H}^+$) and (c) colored
($\tilde {g}$, $\tilde{d}$, $\tilde{u}$, $\tilde{q}$, $\tilde {b}_R$,
$\tilde{t} _R$, $\tilde{q}_3$). In each case the sum over the relevant
LSPs is taken for the results in the Tables. For the $R$-parity
conserving cases these three LSP scenarios are presented in
\TABLEs~\ref{tab:res-rpc-neutral}-\ref{tab:res-rpc-color},
respectively.  We list in each Table how often a specific signature
appears, classified by the number of charged leptons, jets and massive
vector bosons, respectively. Thus in Table~\ref{tab:res-rpc-neutral},
we show in the top row 0, 1 or 2 bosons in the event, $n_v=0,1,2$,
dividing the Table into three big columns. In the second row we
distinguish the number of jets, $n_j$, dividing each of the big
columns into 3 smaller columns.  The first column distinguishes the
number of charged leptons, $n_l$, in the event, yielding five further
rows. Since we are considering the initial production of a colored
particle for each decay chain, we always have at least one jet. Thus
we do not consider $n_j=0$. For a neutral LSP we see that for one
boson, $n_v=1$, two jets, $n_j=2 $, and three charged leptons,
$n_l=3$, we have $39\,029\,940$ hierarchies and dominant decay
scenarios for a given cascade. This corresponds to a relative fraction
of $\sim 0.004\%$, which we have entered in the Table.

In Tables \ref{tab:res-rpc-charged} and \ref{tab:res-rpc-color}, we
have in addition distinguished between signatures without \MET\ (upper
part of each cell) and with \MET\ (lower part of the cell). In Table
\ref{tab:res-rpc-neutral}, we have \MET\ in each case due to the 
neutral and stable LSP.

We can see from the numbers in
\TABLEs~\ref{tab:res-rpc-neutral}-\ref{tab:res-rpc-color} that in the
$R$pC case many hierarchies lead to monojet signatures with and
without \MET\ and without any charged lepton or massive boson. The
transverse momentum is balanced by the charged LSP. The case without
\MET, without a charged lepton, but with one vector boson is unique
for a charged, colorless LSP and also does not appear in any of the
$R$pV scenarios, as can be seen below.  There are also other scenarios
which are specific to $R$-parity conservation. In some cases it is
also possible to determine the kind of LSP. A collection of
distinctive signatures is given in \TABLE~\ref{tab:unique}. Note that
these signatures are expected to be unique if no other signal for SUSY
is discovered: we have focused here on the dominant and best visible
signatures but, of course, sub-dominant signatures of other scenarios
can lead to the same signal. However, in that case it is most likely
that the corresponding \textit{dominat and best visible} signature of
the other sceneario is detected first.
Looking at the case of a neutral LSP given in
\TABLE~\ref{tab:res-rpc-neutral}, we see that monojet events can at
most be accompanied by four charged leptons as already been pointed
out in \cite{Konar:2010bi}. This means that our modified treatment of
third generation particles does not affect this result. The main
reason is that we have treated $\tau$s as jets. Chains like
$\tilde{g}\to\tilde{t}_R\to\tilde{H}^0\to\tilde{\tau}_R\to\tilde
{B}\to\tilde{e}_R\to\tilde{l}\to\tilde{\nu}_\tau$ lead to 4 charged
leptons, 2 jets, 2 $\tau$s and \MET .  And we treat the $\tau$s as
additional jets. One example for a decay chain with $n_l = 4$ and $n_j
=1$ is
\begin{equation}
\label{eq:4jetEvent}
 \tilde{d} \to \tilde{e} \to \tilde{l} \to \tilde{W}^0 \, .
\end{equation}
All three decay channels of the $SU(2)_L$ singlet $\tilde d$ into
$\tilde{l}$, $\tilde{e}$ or $\tilde{W}^0$ are equally suppressed (see
Table~\ref{tab:signatures_colored}) and have therefore be considered.
Obviously, the transition $\tilde{e} \to \tilde{l}$ which emits two
charged leptons is important for this signature. It is much easier to
obtain with a wino instead of a bino LSP. For the latter, $\tilde{e}$
would directly decay to the LSP, as would the $\tilde d$.  Hence, the
observation of this signature would support theories which lead to a
wino LSP, like AMSB.

Surprisingly in \TABLE~\ref{tab:res-rpc-neutral} there are more
hierarchies which lead to monojet events together with four than with
three leptons for a neutral LSP and in the absence of additional
massive bosons. This is related to the right-handedness of the
produced squark. $\tilde{d}/\tilde{u}\to\tilde {e}$ will always lead
to a jet and a charged lepton because the transition happens
dominantly due to an off-shell bino and not a charged wino or
higgsino.  This changes when one looks at two or more jets because
chains like $\tilde{d}\to\tilde{e}\to\tilde{\tau}_R\to\tilde
{H}^+\to\tilde{\nu}$ are possible.
\begin{table}[!bt]
\centering
\begin{tabular}{| c | c | c | c | c | c | c | c | c | c |} 
\hline 
& \multicolumn{3}{c|}{$n_v=0$} & \multicolumn{3}{c|}{$n_v=1$} & \multicolumn{3}{c|}{$n_v=2$} \\ \hline 
$n_l$ & $n_j=1$ & $n_j=2$ & $n_j>2$ & $n_j=1$ & $n_j=2$ & $n_j>2$& $n_j=1$ & $n_j=2$ & $n_j>2$ \\ \hline 
\multirow{2}{*}{0} & $6.45$ & $14.17$ & $13.98$ & $1.77$ & $0.79$ & $0.91$ & $0.13$ & $0.07$ & $0.03$\\ 
 & $3.88$ & $4.35$ & $4.5$ & $0.75$ & $0.31$ & $0.37$ & $0.05$ & $0.01$ & $0.01$\\ 
\hline 
\multirow{2}{*}{1} & $10.32$ & $3.33$ & $7.36$ & $0.22$ & $0.1$ & $0.46$ & $0.04$ & $8\cdot 10^{-3}$ & $0.03$\\ 
 & $3.2$ & $2.19$ & $3.64$ & $0.16$ & $0.17$ & $0.23$ & $4\cdot 10^{-3}$ & $0.01$ & $0.01$\\ 
\hline 
\multirow{2}{*}{2} & $0.53$ & $2.7$ & $3.78$ & $0.09$ & $0.03$ & $0.16$ & $0$ & $6\cdot 10^{-3}$ & $8\cdot 10^{-3}$\\ 
 & $0.55$ & $0.93$ & $1.74$ & $0.05$ & $0.03$ & $0.09$ & $2\cdot 10^{-3}$ & $1\cdot 10^{-3}$ & $5\cdot 10^{-3}$\\ 
\hline 
\multirow{2}{*}{3} & $1.21$ & $0.28$ & $1.49$ & $0.02$ & $8\cdot 10^{-3}$ & $0.06$ & $2\cdot 10^{-3}$ & $5\cdot 10^{-4}$ & $3\cdot 10^{-3}$\\ 
 & $0.37$ & $0.26$ & $0.84$ & $7\cdot 10^{-3}$ & $0.02$ & $0.04$ & $5\cdot 10^{-4}$ & $9\cdot 10^{-4}$ & $2\cdot 10^{-3}$\\ 
\hline 
\multirow{2}{*}{4} & $0$ & $0.2$ & $0.34$ & $1\cdot 10^{-3}$ & $1\cdot 10^{-3}$ & $0.01$ & $0$ & $4\cdot 10^{-4}$ & $8\cdot 10^{-4}$\\ 
 & $0$ & $0.03$ & $0.09$ & $3\cdot 10^{-4}$ & $7\cdot 10^{-5}$ & $3\cdot 10^{-3}$ & $0$ & $4\cdot 10^{-5}$ & $2\cdot 10^{-4}$\\ 
\hline 
\end{tabular}
\caption{Result for $R$-parity conservation and a charged, colorless
  LSP. The notation is as in Table~\ref{tab:res-rpc-neutral}.  The
  upper entry in a given cell of the Table refers to no \MET\, the
  lower entry to \MET\ also being present. All numbers in this table
  refer to percentages of a specific signature.}
\label{tab:res-rpc-charged}
\end{table}

If we compare these results with \TABLE~\ref{tab:res-rpc-charged}
which contains the number of hierarchies with a charged LSP, one can
see that monojet events and four leptons are only possible in case of
exactly one massive boson. Possible hierarchies for these signatures
can easily be derived from \EQ~(\ref{eq:4jetEvent}) by adding a stau
or a charged Higgsino to the end of the cascade:
\begin{align}
 \tilde{d} \to \tilde{e} \to \tilde{l} \to \tilde{W}^0  \to \tilde{\tau}_L \,\, &: \,\, 4 l + 2 j \\
 \tilde{d} \to \tilde{e} \to \tilde{l} \to \tilde{W}^0  \to \tilde{H}^+ \,\, &: \,\, 4 l +  j + v
\end{align}
One might wonder a bit about the zero entry for the $2l+2v+1j
$-signature (no \MET) in \TABLE~\ref{tab:res-rpc-charged}. Two bosons
come from a $\tilde{B}-\tilde{H}-\tilde{W}$ transition, and therefore
$n_v = 2$ events are based on
\begin{equation}
 C \to \tilde{B} \to \tilde{H} \to \tilde{W}^0/\tilde{W}^+ \to \tilde{l}/\tilde{\nu} \to \tilde{e}
\end{equation}
If $C$ is a squark, this will lead to one jet events. All
possible combinations of the last three particles always involve three
external lepton doublets because of lepton number conservation, so
only the combinations $l+\slashed{E}_T$, $2l +\slashed{E}_T $ and $3l$
are possible, but not just $2l$. If one allows more jets the picture
changes because $\tilde{\tau}_R$ can be inserted between
$\tilde{l}/\tilde{\nu} \to \tilde{e}$ leading to the possibility of
two charged leptons with \MET. Again, the special properties of the
emitted $\tau$s play an important role.

Another important feature of a charged LSP is that it provides also
three other signatures beside the four lepton monojet events which can
neither be reached by the other $R$pC cases nor by $R$pV scenarios:
$(n_v, n_j, n_l) = (0,1,2),$ $(1,1,0)$ and $(2,1,0)$.  Possible
cascades to obtain these signatures are\footnote{In this decay chain,
the only unsuppressed $\tilde d$ decay is to a bino. Here the bino is
heavier and thus this becomes a suppressed 3-body decay. Therefore in
Eq.~(\ref{chain1}) the decay $\tilde d\rightarrow\tilde \nu$ with
$1j,\,1l$ is allowed.}:
\begin{align}
 (0,1,2) \, &: \, \, \tilde{d} \to \tilde{\nu} \to \tilde{W}^+ \label{chain1}\\
 (1,1,0) \, &: \, \, \tilde{q} \to \tilde{B} \to \tilde{H}^+ \\
 (2,1,0) \, &: \, \,  \tilde{q} \to \tilde{B} \to \tilde{H}^+ \to \tilde{W}^+ 
\end{align}

\begin{table}[!bt]
\centering
\begin{tabular}{| c | c | c | c | c | c | c | c | c | c | c |} 
\hline 
& \multicolumn{4}{c|}{$n_v=0$} & \multicolumn{3}{c|}{$n_v=1$} & \multicolumn{3}{c|}{$n_v=2$} \\ \hline 
$n_l$ & $n_j=0$ & $n_j=1$ & $n_j=2$ & $n_j>2$ & $n_j \le 1$ & $n_j=2$ & $n_j>2$& $n_j \le 1$ & $n_j=2$ & $n_j>2$ \\ \hline 
\multirow{2}{*}{0} & $35.75$ & $6.7$ & $19.87$ & $12.63$ & $0$ & $1.35$ & $0.85$ & $0$ & $0.11$ & $0.04$\\ 
 & $0$ & $0$ & $1.58$ & $4.02$ & $0$ & $0.03$ & $0.37$ & $0$ & $6\cdot 10^{-3}$ & $0.02$\\ 
\hline 
\multirow{2}{*}{1} & $0$ & $0$ & $0$ & $0$ & $0$ & $0$ & $0$ & $0$ & $0$ & $0$\\ 
 & $0$ & $0$ & $1.05$ & $1.44$ & $0$ & $0.08$ & $0.09$ & $0$ & $0.01$ & $6\cdot 10^{-3}$\\ 
\hline 
\multirow{2}{*}{2} & $0$ & $0$ & $5.1$ & $4.33$ & $0$ & $0.08$ & $0.19$ & $0$ & $0.01$ & $0.01$\\ 
 & $0$ & $0$ & $0.55$ & $1.55$ & $0$ & $7\cdot 10^{-3}$ & $0.09$ & $0$ & $2\cdot 10^{-3}$ & $5\cdot 10^{-3}$\\ 
\hline 
\multirow{2}{*}{3} & $0$ & $0$ & $0$ & $0$ & $0$ & $0$ & $0$ & $0$ & $0$ & $0$\\ 
 & $0$ & $0$ & $0.41$ & $0.47$ & $0$ & $7\cdot 10^{-3}$ & $0.02$ & $0$ & $1\cdot 10^{-3}$ & $1\cdot 10^{-3}$\\ 
\hline 
\multirow{2}{*}{4} & $0$ & $0$ & $0.39$ & $0.51$ & $0$ & $3\cdot 10^{-3}$ & $0.01$ & $0$ & $8\cdot 10^{-4}$ & $1\cdot 10^{-3}$\\ 
 & $0$ & $0$ & $0.06$ & $0.17$ & $0$ & $2\cdot 10^{-4}$ & $8\cdot 10^{-3}$ & $0$ & $1\cdot 10^{-4}$ & $4\cdot 10^{-4}$\\ 
\hline 
\end{tabular}
\caption{Result for $R$-parity conservation and a colored LSP. The
  notation is as in Table~\ref{tab:res-rpc-neutral}.  The upper entry
  in a given cell of the Table refers to no \MET\, the lower entry to
  \MET\ also being present. All numbers in this table refer to
  percentages of a specific signature.}
\label{tab:res-rpc-color}
\end{table}

Next, we consider a colored LSP in the $R$pC case, \textit{ i.e.} a
gluino or a squark. These particles might form `R-hadrons' and come to
rest in the calorimeters. Searches for these signals have been
performed at LEP, the Tevatron
\cite{Heister:2003hc,Arvanitaki:2005nq,Abazov:2007ht} and the LHC
\cite{Aad:2011yf}. If the LSP is the gluino or a squark of the first
two generations it will be produced and will not decay. This leads to
the large number in the $n_j=n_l=n_v=0$ entry because they come with a
huge combinatorial factor of $13!$ taking all possible hierarchies of
the heavier particles into account. The events with one jet but
nothing else are caused by a squark of the third generation as LSP and
a produced gluino.  As soon as one lepton or one massive boson is
involved there have to be at least two jets: the produced colored
particle at the beginning of the cascade as well as the stable colored
particle at the end have to interact with non-colored
particles. Because of baryon number conservation at each vertex at
least two jets will appear. The reason that all events with one or
three charged leptons will also include neutrinos is, of course,
lepton number conservation. However, it is interesting to see that
this also holds when considering the emitted $\tau$s as jets.

\begin{table}[!bt]
\begin{tabular}{| c | c | c | c | c | c | c | c | c | c |} 
\hline 
& \multicolumn{3}{c|}{$n_v=0$} & \multicolumn{3}{c|}{$n_v=1$} & \multicolumn{3}{c|}{$n_v=2$} \\ \hline 
$n_l$ & $n_j=1$ & $n_j=2$ & $n_j>2$ & $n_j=1$ & $n_j=2$ & $n_j>2$& $n_j=1$ & $n_j=2$ & $n_j>2$ \\ \hline 
\multirow{2}{*}{0} & $R$ & & & $c$ & $R$ & & $c$ & $R$ & \\
 & & & & & & & $R$  & & \\
\hline 
\multirow{2}{*}{1} & & & & & & & & & \\
 & $R$ & & & & & &  & & \\
\hline 
\multirow{2}{*}{2} & $c$ & & & \o{} & & &  \o{} & & \\
 & & & & & & &  & & \\
\hline 
\multirow{2}{*}{3} & & & &  & & & & & \\
 & & & & & & & $R$ & & \\
\hline 
\multirow{2}{*}{4} & \o{} & & & $c$ &  & & \o{} &  & \\
 &  & & &  & & &   &  & \\
\hline 
\multirow{2}{*}{5} & $\lambda$ & $\lambda$ & $\slashed{R}$ & $\slashed{R}$ & $\slashed{R}$ &  $\slashed{R}$ & $\slashed{R}$ & $\slashed{R}$ &  $\slashed{R}$ \\
 & $\lambda$ & $\lambda$ &  $\slashed{R}$ & $\slashed{R}$ &  $\slashed{R}$ &  $\slashed{R}$ &$\lambda$  & $\slashed{R}$ &  $\slashed{R}$ \\
\hline 
\multirow{2}{*}{6} & \o{} & \o{} & $\lambda$ & \o{} & $\lambda$ & $\lambda$ & \o{} & \o{} & $\lambda$ \\
 & $\lambda$ & $\lambda$ & $\lambda$ & $\lambda$ & $\lambda$ & $\lambda$ &  $\lambda$ & $\lambda$ & $\lambda$ \\
\hline 
\multirow{2}{*}{7} & \o{} & \o{} & $\lambda$ & $\lambda$ &$\lambda$ & $\lambda$ & \o{} & \o{} & \o{} \\
 & \o{} & $\lambda$ & $\lambda$ & \o{} & \o{} & $\lambda$ & \o{}  & \o{} & \o{} \\
\hline 
\end{tabular} 
\caption{The \textit{dominant and best visible signatures} which
  appear only in a given scenario. We used here $R$ for $R$-parity
  conservation in general and $n$ (neutral), $c$ (charged), $h$
  (colored) if also the kind of LSP can be determined. When a specific
  signal is just possible $R$-parity violation, we used $\slashed{R}$,
  while we give the corresponding coupling this is only possible for
  one scenario. In addition, we use \o{} for cases which are neither
  covered in the $R$pV nor $R$pC case and therefore unique in that
  sense that they demand an extension of the MSSM. The upper line in a
  cell is for the case without \MET, the lower part with \MET. Note,
  this table does not include a column for $n_j=0$, but we point out
  that $n_j=0$, $n_l=0$, $n_\nu = 1$ without \MET\ is unique for a
  colored LSP in the $R$pC case.}
\label{tab:unique}
\end{table}

We want to give here one example for a decay chain with the maximal
amount of charged leptons and massive bosons but the minimal amount of
jets: four leptons together with two massive bosons and two jets appear
for instance in
\begin{equation}
\label{eq:4jetColored}
\tilde{q} \to \tilde{B} \to \tilde{H} \to \tilde{W}^0 \to \tilde{l} \to \tilde{e} \to \tilde{t}_R
\end{equation}
The produced squark has not necessarily to be left-handed and the LSP
can also be left-handed or a sbottom.
 
In compiling the $R$pC scenarios we first compared with
Ref.~\cite{Konar:2010bi} and found agreement. The results presented
above differ, since we have treated the third generation leptons
separately and have also treated the two leptons in the $SU(2)_L$
doublets separately.

\subsubsection*{Summary for $R$pC}
We have distinguished three $R$pC cases: (i) for a neutral LSP always
\MET\ is present and monojet events can be accompanied with at most
four charged leptons.  Interestingly, there are more hierarchies
leading to four than to three charged leptons and one jet. (ii) In the
case of a charged and colorless LSP monojet events with four charged
leptons can only appear for $n_v = 1$. The $n_j = n_v = 1, n_l =4$
with \MET\ is unique and can neither appear in any other $R$pV nor
$R$pC scenario. In addition, there are three other unique signatures
for a charged, coloreless LSP: $(n_v,n_j,n_l) = (0, 1, 2), (1, 1, 0)$
and $(2, 1, 0)$. (iii) The most often appearing hierarchies for a
colored LSP are either $n_j = n_v = n_l = 0$ if the LSP a gluino or a
squark of the first two generations. For a squark of the third
generation there are only signatures with at least two jets. For all
three cases it turned out that the treatment of the $\tau$s is
crucial: the results of Ref.~\cite{Konar:2010bi}, which did not differ
between the third and the other two generations, are only recovered
because $\tau$s are counted as jets.

\subsection{Bilinear $R$pV}

\begin{table}[!bt]
\centering
\begin{tabular}{| c | c | c | c | c | c | c | c | c | c |} 
\hline 
& \multicolumn{3}{c|}{$n_v=0$} & \multicolumn{3}{c|}{$n_v=1$} & \multicolumn{3}{c|}{$n_v=2$} \\ \hline 
$n_l$ & $n_j=1$ & $n_j=2$ & $n_j>2$ & $n_j=1$ & $n_j=2$ & $n_j>2$& $n_j=1$ & $n_j=2$ & $n_j>2$ \\ \hline 
\multirow{2}{*}{0} & $0$ & $0$ & $4.25$ & $0$ & $0$ & $0.27$ & $0$ & $0$ & $0.02$\\ 
 & $2.99$ & $0.75$ & $15.72$ & $2.24$ & $0.75$ & $1.02$ & $0$ & $0.06$ & $0.09$\\ 
\hline 
\multirow{2}{*}{1} & $5.98$ & $4.49$ & $15.28$ & $7.28$ & $2.73$ & $4.73$ & $1.99$ & $0.49$ & $0.43$\\ 
 & $0$ & $0$ & $3.42$ & $0.8$ & $0.91$ & $0.9$ & $0.04$ & $0.17$ & $0.04$\\ 
\hline 
\multirow{2}{*}{2} & $0$ & $0$ & $1.29$ & $0$ & $0$ & $0.03$ & $0$ & $0$ & $3\cdot 10^{-3}$\\ 
 & $2.17$ & $0.7$ & $6.31$ & $1.11$ & $0.3$ & $0.63$ & $0.14$ & $0.02$ & $0.03$\\ 
\hline 
\multirow{2}{*}{3} & $0$ & $0$ & $3.16$ & $1.38$ & $0.24$ & $0.98$ & $0.09$ & $0.01$ & $0.05$\\ 
 & $0$ & $0$ & $0.95$ & $0.23$ & $0.18$ & $0.22$ & $7\cdot 10^{-3}$ & $0.02$ & $6\cdot 10^{-3}$\\ 
\hline 
\multirow{2}{*}{4} & $0$ & $0$ & $0.12$ & $0$ & $0$ & $3\cdot 10^{-3}$ & $0$ & $0$ & $3\cdot 10^{-4}$\\ 
 & $0.17$ & $0.04$ & $0.78$ & $0.1$ & $0.03$ & $0.1$ & $5\cdot 10^{-3}$ & $4\cdot 10^{-4}$ & $2\cdot 10^{-3}$\\ 
\hline 
\multirow{2}{*}{5} & $0$ & $0$ & $0.2$ & $0.03$ & $3\cdot 10^{-3}$ & $0.03$ & $1\cdot 10^{-3}$ & $7\cdot 10^{-5}$ & $6\cdot 10^{-4}$\\ 
 & $0$ & $0$ & $0.05$ & $4\cdot 10^{-3}$ & $8\cdot 10^{-3}$ & $7\cdot 10^{-3}$ & $0$ & $2\cdot 10^{-5}$ & $1\cdot 10^{-4}$\\ 
\hline 
\end{tabular} 
\caption{Results for bilinear $R$-parity violation. The notation is as
  in Table~\ref{tab:res-rpc-neutral}.  The upper entry in a given cell
  of the Table refers to no \MET\, the lower entry to \MET\ also being
  present. All numbers in this table refer to percentages of a
  specific signature.}
\label{res-rpv-epsilon}
\end{table}

We turn now to the results for bilinear $R$pV given in
\TABLE~\ref{res-rpv-epsilon}\footnote{This also covers models which
  contain the bilinear model when integrating out heavy additional
  states, e.g.~spontaneous $R$-parity violation
  \cite{Masiero:1990uj,Hirsch:2008ur} or the $\mu\nu$SSM
  \cite{LopezFogliani:2005yw,Bartl:2009an}.}. Since the LSP decays, we
present in each entry in the table the sum of the hierarchies for all
possible LSPs. Here, the maximal number of charged lepton tracks is
five. This is more than in the $R$pC case. Five charged leptons can be
reached for instance due to the decay channels listed in
\TABLE~\ref{tab:RpVmodes} and the signatures given in
\EQ~(\ref{eq:4jetEvent}) for a wino LSP or \EQ~(\ref{eq:4jetColored})
for a squark LSP.

There are two other interesting results: \textit{(i)} the signal $v +
j + l$ ($n_v = n_j = n_l = 1$) without \MET\ can only be generated in
bilinear $R$pV with a general LSP as well as in $R$pC with a charged
LSP. Thus if this signature is observed and if the LSP is found and
uncharged, bilinear $R $-parity violation is the best candidate to
explain this within supersymmetry. \textit{(ii)} Monojet events with
\MET\ and zero or one massive boson appear much more often in bilinear 
$R$pV than in the other cases.

In \TABLE~\ref{res-rpv-epsilon}, the origin of the $n_v=n_j=n_l=1$
entries without \MET\ are the hierarchies with a wino LSP. As long as
the produced squark is left-handed it will always decay dominantly
into the LSP independently of the mass ordering of the other
states. However, if the produced squark is right-handed then for this
signature ($n_v=n_j=n_l=1$) the bino cannot be lighter because then the
squark would not decay directly to the LSP.

The large number of monojet events with \MET\ and one charged lepton
comes from the direct
decays of right-handed squarks. Therefore, its amount is exactly half
of the $n_v = n_j = n_l = 0$-entry in
\TABLE~\ref{tab:res-rpc-color}.

The zero entries in \TABLE~\ref{res-rpv-epsilon} indicate that there
are many scenarios which can disfavor bilinear $R$pV. For instance, an
even number of charged leptons is only (dominantly) possible
together with \MET\ as long as less than three jets are present. The
reason is again the special properties of the $\tau$, \textit{e.g.}
\begin{equation}
\tilde{q} \to \tilde{e} \to \tilde{l} \to \tilde{\tau}_L 
\end{equation}
leads to 4 jets and 4 charged leptons.Two jets arise from the
$\tilde\tau$-LSP decay.

\begin{table}[!bt]
\centering
\begin{tabular}{| c | c | c | c | c | c | c | c | c | c |} 
\hline 
& \multicolumn{3}{c|}{$n_v=0$} & \multicolumn{3}{c|}{$n_v=1$} & \multicolumn{3}{c|}{$n_v=2$} \\ \hline 
$n_l$ & $n_j=1$ & $n_j=2$ & $n_j>2$ & $n_j=1$ & $n_j=2$ & $n_j>2$& $n_j=1$ & $n_j=2$ & $n_j>2$ \\ \hline 
\multirow{2}{*}{1} & $0$ & $0$ & $0$ & $0$ & $0$ & $0$ & $0$ & $0$ & $0$\\ 
 & $0$ & $4.28$ & $4.15$ & $0$ & $0.17$ & $0.22$ & $0$ & $0.02$ & $0.01$\\ 
\hline 
\multirow{2}{*}{2} & $0$ & $0.4$ & $0.31$ & $0$ & $0.08$ & $0.03$ & $0$ & $5\cdot 10^{-3}$ & $1\cdot 10^{-3}$\\ 
 & $24.17$ & $10.63$ & $20.16$ & $1.32$ & $0.45$ & $1.4$ & $0.15$ & $0.03$ & $0.07$\\ 
\hline 
\multirow{2}{*}{3} & $5.98$ & $2.18$ & $3.32$ & $1.13$ & $0.28$ & $0.29$ & $0.09$ & $0.02$ & $4\cdot 10^{-3}$\\ 
 & $0.39$ & $2.25$ & $2.21$ & $0.04$ & $0.12$ & $0.1$ & $4\cdot 10^{-4}$ & $4\cdot 10^{-3}$ & $5\cdot 10^{-3}$\\ 
\hline 
\multirow{2}{*}{4} & $0$ & $0.15$ & $0.14$ & $0$ & $4\cdot 10^{-3}$ & $6\cdot 10^{-3}$ & $0$ & $3\cdot 10^{-4}$ & $3\cdot 10^{-4}$\\ 
 & $3.3$ & $0.74$ & $5.82$ & $0.15$ & $0.03$ & $0.2$ & $3\cdot 10^{-3}$ & $6\cdot 10^{-4}$ & $0.01$\\ 
\hline 
\multirow{2}{*}{5} & $0.58$ & $0.1$ & $0.68$ & $0.05$ & $5\cdot 10^{-3}$ & $0.03$ & $6\cdot 10^{-5}$ & $1\cdot 10^{-5}$ & $3\cdot 10^{-4}$\\ 
 & $0.11$ & $0.31$ & $0.38$ & $4\cdot 10^{-3}$ & $0.01$ & $0.01$ & $8\cdot 10^{-5}$ & $3\cdot 10^{-4}$ & $7\cdot 10^{-4}$\\ 
\hline 
\multirow{2}{*}{6} & $0$ & $0$ & $0.01$ & $0$ & $1\cdot 10^{-4}$ & $2\cdot 10^{-4}$ & $0$ & $0$ & $2\cdot 10^{-5}$\\ 
 & $0.15$ & $0.03$ & $0.46$ & $4\cdot 10^{-3}$ & $6\cdot 10^{-4}$ & $9\cdot 10^{-3}$ & $9\cdot 10^{-5}$ & $1\cdot 10^{-5}$ & $8\cdot 10^{-4}$\\ 
\hline 
\multirow{2}{*}{7} & $0$ & $0$ & $0.01$ & $7\cdot 10^{-4}$ & $2\cdot 10^{-5}$ & $2\cdot 10^{-5}$ & $0$ & $0$ & $0$\\ 
 & $0$ & $6\cdot 10^{-3}$ & $2\cdot 10^{-3}$ & $0$ & $0$ & $5\cdot 10^{-6}$ & $0$ & $0$ & $0$\\ 
\hline 
\end{tabular} 
\caption{Results for $R$-parity violation: $\lambda$ term. The
  notation is as in Table~\ref{tab:res-rpc-neutral}.  The upper entry
  in a given cell of the Table refers to no \MET\, the lower entry to
  \MET\ also being present. All numbers in this table refer to
  percentages of a specific signature.}
\label{res-rpv-lam}
\end{table}

\subsubsection*{Summary for bilinear $R$pV}
Bilinear $R$pV can often be disfavored because many signatures are
only sub-dominant. This is for instance the case for a even number of
charged leptons and \MET\ and less than three jets.  However, monojet
events with \MET\ and less than two massive vector bosons appear much
more often in bilinear $R$pV than in all other cases. Furthermore,
signatures with one jet, one charged lepton, one massive vector boson
and without \MET\ are only possible for bilinear $R$pV or $R$pC with a
colored LSP.

\subsection{Trilinear $R$pV}

Next we consider the trilinear $R$pV scenario $\lambda$\LLE. 
This interaction leads to an increase in the number of charged
leptons in the final states. Up to seven lepton tracks with and
without \MET\ are possible, \textit{c.f.}
\TABLE~\ref{res-rpv-lam}. This is found for the mass hierarchies
\begin{equation}
\label{eq:7lEvent}
\tilde{q} \to \tilde{B} \to \tilde{e} \to \tilde{l} \to \tilde{\nu}_\tau \to \tilde{H}^\pm
\end{equation}
thanks to the three leptons in the final decay of the LSP, $\tilde H^
\pm \to l^\pm l^+ l^-$ (the $\tilde l\rightarrow\tilde\nu_\tau$ should
go through a neutral wino). However, even in the unconstrained version
of the MSSM it is hard to obtain a chargino LSP, consistent with the
LEP constraints. Therefore, this signature might not only point to
$R$-parity violation but also to some other extension of the MSSM. In
contrast, events with 6 charged leptons, which are still outstanding,
are possible for a neutral LSP as well as for a stop LSP, which is
possible within $R$pV mSUGRA \cite{Dreiner:2009fi}. A potential
spectrum hierarchy leading to 6 leptons is
\begin{equation}
\tilde{q} \to \tilde{B} \to \tilde{e} \to \tilde{l} \to \tilde{H}^0/
\tilde{t}_L 
\end{equation}
Of course, also the cascades in \EQs~(\ref{eq:4jetEvent}) and
(\ref{eq:4jetColored}) do the job. 

The reason that there are no seven jets events possible with two
massive bosons is that the bino and the wino have to be present and
have to couple to the higgsino. Therefore, the transitions $\tilde{C} -
\tilde{B}/\tilde{W} -\tilde{l}/\tilde{e}_R$ are not possible. Another
remarkable signature is the one with seven charged leptons but only
one jet and one massive boson. This signature can be obtained by
replacing $\tilde{\nu}_\tau$ by $\tilde{W}^0$ in 
\EQ~(\ref{eq:7lEvent}).

On the other hand, the $\lambda$-interaction can be easily disfavored
by the observation of supersymmetric events with less than two
leptons. The only possibility to get one-lepton events are due to the
decay of $\tau$s, for example in the chains
\begin{equation}
\tilde{q} \to \tilde{\tau}_L \,, \hspace{1cm} \tilde{q} \to \tilde{B}  
\to \tilde{H}^0 \to \tilde{\tau}_L \,, \hspace{1cm} \tilde{q} \to 
\tilde{B}  \to \tilde{H}^0 \to \tilde{W}^0 \to \tilde{\tau}_L
\end{equation}
with two jets and zero, one or two massive vector bosons.

\begin{table}[!bt]
\centering
\begin{tabular}{| c | c | c | c | c | c | c | c | c | c |} 
\hline 
& \multicolumn{3}{c|}{$n_v=0$} & \multicolumn{3}{c|}{$n_v=1$} & \multicolumn{3}{c|}{$n_v=2$} \\ \hline 
$n_l$ & $n_j=1$ & $n_j=2$ & $n_j>2$ & $n_j=1$ & $n_j=2$ & $n_j>2$& $n_j=1$ & $n_j=2$ & $n_j>2$ \\ \hline 
\multirow{2}{*}{0} & $0$ & $0$ & $4.17$ & $0$ & $0$ & $0.26$ & $0$ & $0$ & $0.02$\\ 
 & $0$ & $0.73$ & $11.41$ & $0$ & $0$ & $0.75$ & $0$ & $0$ & $0.06$\\ 
\hline 
\multirow{2}{*}{1} & $5.87$ & $3.67$ & $35.35$ & $0$ & $0$ & $3.64$ & $0$ & $0$ & $0.26$\\ 
 & $0$ & $0$ & $9.75$ & $0$ & $0$ & $0.65$ & $0$ & $0$ & $0.03$\\ 
\hline 
\multirow{2}{*}{2} & $0$ & $0$ & $1.27$ & $0$ & $0$ & $0.03$ & $0$ & $0$ & $3\cdot 10^{-3}$\\ 
 & $0$ & $0$ & $5.66$ & $0$ & $0$ & $0.38$ & $0$ & $0$ & $0.01$\\ 
\hline 
\multirow{2}{*}{3} & $0$ & $0$ & $10.05$ & $0$ & $0$ & $0.4$ & $0$ & $0$ & $0.02$\\ 
 & $0$ & $0$ & $3.24$ & $0$ & $0$ & $0.15$ & $0$ & $0$ & $9\cdot 10^{-3}$\\ 
\hline 
\multirow{2}{*}{4} & $0$ & $0$ & $0.11$ & $0$ & $0$ & $3\cdot 10^{-3}$ & $0$ & $0$ & $2\cdot 10^{-4}$\\ 
 & $0$ & $0$ & $0.99$ & $0$ & $0$ & $0.03$ & $0$ & $0$ & $3\cdot 10^{-3}$\\ 
\hline 
\multirow{2}{*}{5} & $0$ & $0$ & $0.71$ & $0$ & $0$ & $0.02$ & $0$ & $0$ & $2\cdot 10^{-3}$\\ 
 & $0$ & $0$ & $0.27$ & $0$ & $0$ & $8\cdot 10^{-3}$ & $0$ & $0$ & $7\cdot 10^{-4}$\\ 
\hline 
\end{tabular} 
\caption{Results for $R$-parity violation: $\lambda'$ term. The
  notation is as in Table~\ref{tab:res-rpc-neutral}.  The upper entry
  in a given cell of the Table refers to no \MET\, the lower entry to
  \MET\ also being present. All numbers in this table refer to
  percentages of a specific signature.}
\label{res-rpv-lamp}
\end{table}

The second lepton number violating coupling, $\lambda'$, usually leads
to multi-jet events. The only exception is the case without any
massive boson and without \MET\ but with just one charged lepton. Such
an event happens if the LSP is a squark of the first two generations
($\tilde{d},\,\tilde u$ or $\tilde{q}$).  In that case, the LSP is
produced and decays directly due to a $R$-parity violating channel
without any intermediate cascade, \textit{e.g.} $\tilde u_L\to
l^+ + d_R$. Note that $\tilde{u}$ does not directly interact via the $lqd$
operator. Thus the decay proceeds via two off-shell particles leading
to three jets and one charged lepton. The two jet events in Table
\ref{res-rpv-lamp} are a consequence of a gluino LSP. On the other
side, theoretically up to eleven jets can be caused by the cascade
\begin{equation}
 \tilde{d} \to \tilde{\tau}_L \to \tilde{W}^0 \to \tilde{b}_L \to \tilde{b}_R \to \tilde{t}_R\,.
\end{equation}
The stop LSP decays into three jets and a charged lepton.

\begin{table}[!bt]
\centering
\begin{tabular}{| c | c | c | c | c | c | c | c | c | c |} 
\hline 
& \multicolumn{3}{c|}{$n_v=0$} & \multicolumn{3}{c|}{$n_v=1$} & \multicolumn{3}{c|}{$n_v=2$} \\ \hline 
$n_l$ & $n_j=2$ & $n_j=3$ & $n_j>3$ & $n_j=2$ & $n_j=3$ & $n_j>3$& $n_j=2$ & $n_j=3$ & $n_j>3$ \\ \hline 
\multirow{2}{*}{0} & $9.38$ & $4.69$ & $37.98$ & $0$ & $0$ & $4.21$ & $0$ & $0$ & $0.3$\\ 
 & $0$ & $0$ & $7.87$ & $0$ & $0$ & $0.76$ & $0$ & $0$ & $0.03$\\ 
\hline 
\multirow{2}{*}{1} & $0$ & $0$ & $0$ & $0$ & $0$ & $0$ & $0$ & $0$ & $0$\\ 
 & $0$ & $0$ & $8.19$ & $0$ & $0$ & $0.56$ & $0$ & $0$ & $0.03$\\ 
\hline 
\multirow{2}{*}{2} & $0$ & $0$ & $17.45$ & $0$ & $0$ & $0.65$ & $0$ & $0$ & $0.05$\\ 
 & $0$ & $0$ & $3.71$ & $0$ & $0$ & $0.22$ & $0$ & $0$ & $0.01$\\ 
\hline 
\multirow{2}{*}{3} & $0$ & $0$ & $0$ & $0$ & $0$ & $0$ & $0$ & $0$ & $0$\\ 
 & $0$ & $0$ & $1.42$ & $0$ & $0$ & $0.06$ & $0$ & $0$ & $3\cdot 10^{-3}$\\ 
\hline 
\multirow{2}{*}{4} & $0$ & $0$ & $1.92$ & $0$ & $0$ & $0.05$ & $0$ & $0$ & $4\cdot 10^{-3}$\\ 
 & $0$ & $0$ & $0.44$ & $0$ & $0$ & $0.02$ & $0$ & $0$ & $1\cdot 10^{-3}$\\ 
\hline 
\end{tabular} 
\caption{Results for $R$-parity violation: $\lambda''$ term. The
  notation is as in Table~\ref{tab:res-rpc-neutral}.  The upper entry
  in a given cell of the Table refers to no \MET\, the lower entry to
  \MET\ also being present. All numbers in this table refer to
  percentages of a specific signature.}
\label{res-rpv-lampp}
\end{table}

The last remaining $R$pV case is the baryon number violating term $
\lambda''$\UDD. This interaction causes a large
number of additional jets and therefore monojet events are not
possible.  Events with exactly two jets can only occur if the LSP
is the directly produced squark. In case of a gluino LSP three jets
are emitted in the $R$pV decay.

As a consequence of this, this scenario would be under pressure by the
observation of events with less then 4 jets accompanied by some
charged leptons or an amount of charged leptons which is larger than
expected already for the $R$pC case. In contrast, multi-jet events can
also appear easily in the $R$-parity conserving case as well as for
lepton-number violating operators. This renders it rather difficult to
find a signature which privileges the $\lambda''$ case, since we have
seen that already in the \LQDspace scenario up to eleven jets are possible.

Finally, we have here treated each final state quark as a separate
jet. This need not be the case. If the jets resulting from a decaying
particle are clumped, they can manifest an interesting substructure
\cite{Butterworth:2008iy}. This is particularly relevant for the \UDDspace
case \cite{Butterworth:2009qa}. See also the recent work in
Ref.~\cite{Allanach:2012vj}.

\begin{table}
\begin{tabular}{| c | c | c | c | c | c | c | c | c | c |} 
\hline 
& \multicolumn{3}{c|}{$n_v=0$} & \multicolumn{3}{c|}{$n_v=1$} & \multicolumn{3}{c|}{$n_v=2$} \\ \hline 
$n_l$ & $n_j=1$ & $n_j=2$ & $n_j>2$ & $n_j=1$ & $n_j=2$ & $n_j>2$& $n_j=1$ & $n_j=2$ & $n_j>2$ \\ \hline 
\multirow{2}{*}{0} & $n$,$\slashed{R}$& $n$,$\epsilon$,$\lambda$,$\lambda'$ & $n$,$\lambda$  & $n$,$h$,$\slashed{R}$ & $n$,$\epsilon$,$\lambda$,$\lambda''$ &$n$,$\lambda$,$\lambda'$ & $n$,$h$,$\epsilon$,$\lambda$,$\lambda''$ & $n$,$\epsilon$,$\lambda'$,$\lambda$,$\lambda'$,$\lambda''$ & $n$,$\lambda$  \\
 & $h$,$\lambda$,$\lambda'$,$\lambda''$ & $\lambda$,$\lambda''$ & $\lambda$  & $h$,$\lambda$,$\lambda'$,$\lambda''$  & $\lambda$,$\lambda'$,$\lambda''$ & $\lambda$ & $h$,$\slashed{R}$ &  $\lambda$,$\lambda'$,$\lambda''$ & $\lambda$  \\
\hline 
\multirow{2}{*}{1} & $n$,$h$,$\lambda$,$\lambda''$  & $n$,$h$,$\lambda$,$\lambda''$  & $n$,$h$,$\lambda$,$\lambda''$  &  $n$,$h$,$\lambda$,$\lambda'$,$\lambda''$ &  $n$,$h$,$\lambda$,$\lambda'$,$\lambda''$ &  $n$,$h$,$\lambda$,$\lambda''$ & $n$,$h$, $\lambda$,$\lambda'$,$\lambda''$ &$n$,$h$, $\lambda$,$\lambda'$,$\lambda''$  & $n$,$h$,$\lambda$,$\lambda''$  \\
 & $h$,$\slashed{R}$ & $\epsilon$,$\lambda'$,$\lambda''$ &  &  $h$,$\lambda$,$\lambda'$,$\lambda''$ &  $\lambda'$,$\lambda''$ &  & $h$,$\lambda$,$\lambda'$,$\lambda''$ & $\lambda'$,$\lambda''$  &   \\
\hline 
\multirow{2}{*}{2} & $n$,$h$,$\slashed{R}$ & $n$,$\epsilon$,$\lambda'$,$\lambda''$ & $n$ & $n$,$h$,$\slashed{R}$ & $n$,$\epsilon$,$\lambda'$,$\lambda''$ & $n$ & all &$n$, $\epsilon$,$\lambda'$,$\lambda''$ & $n$   \\
 & $h$,$\lambda'$,$\lambda''$ & $\lambda'$,$\lambda''$  &  &  $h$,$\lambda'$,$\lambda''$ & $\lambda'$,$\lambda''$  &  & $h$,$\lambda'$,$\lambda''$  & $\lambda'$,$\lambda''$  &   \\
\hline 
\multirow{2}{*}{3} & $n$,$h$,$\epsilon$,$\lambda'$,$\lambda''$ & $n$,$h$,$\epsilon$,$\lambda'$,$\lambda''$ & $n$,$h$,$\lambda''$ & $n$,$h$,$\lambda'$,$\lambda''$ & $n$,$h$,$\lambda'$,$\lambda''$  & $n$,$h$,$\lambda''$ & $n$,$h$, $\lambda'$,$\lambda''$ & $n$,$h$,$\lambda'$,$\lambda''$ & $n$,$h$,$\lambda''$   \\
 & $h$,$\epsilon$,$\lambda'$,$\lambda''$ & $\epsilon$,$\lambda'$,$\lambda''$ &  & $h$,$\lambda'$,$\lambda''$ & $\lambda'$,$\lambda''$ &  & $h$,$\slashed{R}$ & $\lambda'$,$\lambda''$  &   \\
\hline 
\multirow{2}{*}{4} & all & $n$,$\epsilon$,$\lambda'$,$\lambda''$ & $n$  & $n$, $h$,$\slashed{R}$ & $c$,$\epsilon$,$\lambda'$,$\lambda''$ & $n$  & all & $n$,$\epsilon$,$\lambda'$,$\lambda''$ & $n$  \\
 & $c$,$h$,$\lambda'$,$\lambda''$ & $\lambda'$,$\lambda''$ &  & $h$,$\lambda'$,$\lambda''$ & $\lambda'$,$\lambda''$ &  & $c$,$h$,$\lambda'$,$\lambda''$ & $\lambda'$,$\lambda''$ & \\ 
\hline 
\multirow{2}{*}{5} & $R$,$\epsilon$,$\lambda'$ & $R$,$\epsilon$,$\lambda'$ & $R$ & $R$,$\lambda'$ & $R$,$\lambda'$ & $R$ & $R$,$\lambda'$ & $R$,$\lambda'$ & $R$\\ 
 & $R$,$\epsilon$,$\lambda'$ & $R$,$\epsilon$,$\lambda'$ & $R$ & $R$,$\lambda'$ & $R$,$\lambda'$ & $R$ & $R$,$\epsilon$,$\lambda'$ & $R$,$\lambda'$ & $R$\\ 
\hline 
\multirow{2}{*}{6} & all & all & $R$,$\epsilon$, $\lambda'$, $\lambda''$ & all & $R$,$\epsilon$, $\lambda'$, $\lambda''$ & $R$,$\epsilon$, $\lambda'$, $\lambda''$ & all & all & $R$,$\epsilon$, $\lambda'$, $\lambda''$\\ 
& $R$,$\epsilon$, $\lambda'$, $\lambda''$ & $R$,$\epsilon$, $\lambda'$, $\lambda''$ & $R$,$\epsilon$, $\lambda'$, $\lambda''$ & $R$,$\epsilon$, $\lambda'$, $\lambda''$ & $R$,$\epsilon$, $\lambda'$, $\lambda''$ & $R$,$\epsilon$, $\lambda'$, $\lambda''$ & $R$,$\epsilon$, $\lambda'$, $\lambda''$ & $R$,$\epsilon$, $\lambda'$, $\lambda''$ & $R$,$\epsilon$, $\lambda'$, $\lambda''$\\ 
\hline 
\multirow{2}{*}{7} & all & all & $R$,$\epsilon$, $\lambda'$, $\lambda''$ & $R$,$\epsilon$, $\lambda'$, $\lambda''$ & $R$,$\epsilon$, $\lambda'$, $\lambda''$ & $R$,$\epsilon$, $\lambda'$, $\lambda''$ & all & all & all\\ 
& all & $R$,$\epsilon$, $\lambda'$, $\lambda''$ & $R$,$\epsilon$, $\lambda'$, $\lambda''$ & all & all & $R$,$\epsilon$, $\lambda'$, $\lambda''$ & all & all & all\\ 
\hline 
\end{tabular} 
\caption{Signatures which do not appear dominantly in a given model.
  We used here $R$ for $R$-parity conservation in general and $n$
  (neutral), $c$ (charged), $h$ (colored) if also the kind of LSP can
  be determined. When a specific signal is just possible $R$-parity
  violation, we used $\slashed{R}$, while we give the corresponding
  coupling this is only possible for one scenario. The upper line in a
  cell is for the case without \MET, the lower part with \MET. Note,
  this table does not include a column for $n_j=0$, but we point out
  that $n_j=0$, $n_l=0$, $n_\nu = 1$ without \MET\ would exclude all
  cases but $h$, while any other combination with $n_j=0$ would
  disfavor all scenarios discussed here.\label{tab:ruledout} }
\end{table}

\subsubsection*{Summary for trilinear $R$pV}
The three cases of trilinear $R$pV have very different features: (i)
in case of $\lambda$\LLE\, signatures with up to seven charged leptons
are possible while in all other setups at most five are
possible. However, $n_l=7$ is only possible for a chargino LSP what is
usually hard to reach in the MSSM, $n_l=6$ works also for a Higgsino
or stau LSP. (ii) If the LSP is not a squark of the first two
generations, and therefore directly produced, $\lambda'$\LQD\ leads to
multi-jet events.  Therefore, it is very hard to distinguish this case
from the one with $R$pC and a colored LSP. The main difference between
both is the possibility of five charged leptons. (iii) Finally,
$\lambda''$\UDD\ might be the scenario which is the hardest one to be
proven correct. The reason is that only in 14\% of all hierarchies it
is possible to get events with less than 4 jets while events with zero
or one jet are not possible at all.

\subsection{Exclusion Summary}

We have collected in Table~\ref{tab:ruledout} all scenarios which can
be disfavored by the observation of a specific signature. Obviously,
especially events with just one or two jets as well as multi-lepton
events are a very effective selection criterion. Thus for $n_v=2$ and
$n_l=7$ we have written `all' in the three boxes for $n_j=1$, $n_j=2$
and $n_j>2$. This means that \textit{no} model predicts this signature
dominantly. In the box for $n_v=0,\,n_j>2$ and $n_l=4$ we have the
entry `$n$' meaning this signature does \textit{not} appear dominantly
if the LSP is neutral.

A word of caution should be added here.  This table applies in case
that one signature is observed at the LHC. However, we emphasize that
other signatures, which do not belong to the set of \textit{dominant
  and best visible signatures} considered here, are possible.
Nevertheless, in those cases it is very likely that the most visible
signature is detected first.

Furthermore, there are also signatures which cannot be reached
dominantly in any of the presented scenarios. Such an observation
would directly point to an extension of the MSSM. It might be a
surprise that not only events with four or more charged leptons are
hard to produce in the MSSM with and without $R$-parity violation, but
also one jet, two charged leptons and two massive bosons is not a
result of the dominant decay chain of any of the $14!$ hierarchies
considered here.

\section{Conclusion}
\label{sec:conclusion}
We have presented in this work the collider signatures appearing in a
very general realization of the MSSM based on 14 unrelated mass
parameters. These mass parameters lead to $14!\approx9\cdot10^{10}$
particle orderings. We go beyond previous work to allow for separate
third generation soft supersymmetry breaking parameters.  For each
mass ordering of the supersymmetric fields, we have determined the
dominant decay modes. We imagine this as a chain of decays giving a
cascade overall. Each supersymmetric particle can decay to all other
lighter supersymmetric particles, fixed by that mass ordering. For
these we consider decays involving the mixing between sparticles to be
subdominant. If a 2-body decay mode exists, then 3-body decays are
subdominant. For all sfermions other than the third generation,
Higgs(ino) couplings are subdominant. We can thus classify all decays
for each supersymmetric particle which are summarized in
\TABLEs~\ref{tab:signatures_colored}-\ref{tab:signatures_fermion}.

These tables are the basis for all our results. Each dominant decay
leads to a specific signature. For our signatures we consider: charged
leptons, jets, massive bosons ($W^\pm, Z$, Higgs) and \MET. We have
then systematically gone through all possible mass ordering and
determined the dominant decay signatures. We have summarized the
results for several supersymmetric models, both $R$-parity conserving
and $R$-parity violating, in a series of tables.

In Table \ref{tab:res-rpc-neutral}, we consider the case of $R$-parity
conservation with a neutralino LSP. This closely resembles the
previous work in \cite{Konar:2010bi}.  We go beyond their work to
consider the third generation parameters as separate.  In the table,
we list the relative frequency with which the various signatures
occur.

In Tables \ref{tab:res-rpc-charged} and \ref{tab:res-rpc-color} we
perform the analogous analysis for a possible electrically charged or
color charged stable LSP. Here we have in addition determined whether
the final state signature includes \MET\ or not.

We found signatures which are specific for a particular $R$pV scenario
or for the $R$pC case. However, it has also been shown that there is
no hierarchy in the case of the \LQD\ couplings which leads to a
signature that cannot be reached by another setup as well. This makes
it a bit difficult to favor one of these possibilities by looking just
at the signatures produced at the LHC. However, not only pure monojet
events are only possible with conserved $R$-parity, but also other
signatures like two jets and one massive vector boson will not appear
dominantly in $R$pV scenarios while they do in $R$pC. In addition,
there is at least one signature which is dominant only in trilinear
$R$pV and many multi-lepton signatures which are outstanding for
\LLE\ couplings. Furthermore, we also found several signatures which
can, at least, highly disfavor some of the scenarios considered here
due to the impossibility to find hierarchies that can produce such a
signal in its dominant decay mode.

\section*{Acknowledgements}
We thank Martin Hirsch for discussions and collaboration in the early
stage of this project. One of us (HD) would like to thank Konstantin
Matchev for a detailed discussion explaing his work in
Ref. \cite{Konar:2010bi}, which in turn triggered this work. AV would
like to thank the Bonn theory group for their hospitality. This work
has been supported in part by the Helmholtz Alliance ``Physics at the
Terascale'' and W.P.\ in part by the DFG, project No. PO-1337/2-1. AV
acknowledges support from the ANR project CPV-LFV-LHC {NT09-508531}.


\begin{thebibliography}{10}

\bibitem{Aad:2009wy}
  G.~Aad {\it et al.}  [The ATLAS Collaboration],
  arXiv:0901.0512 [hep-ex].

\bibitem{Chatrchyan:2008zzk}
  S.~Chatrchyan {\it et al.}  [CMS Collaboration],
  JINST {\bf 3} (2008) S08004.

\bibitem{Martin:1997ns}
  S.~P.~Martin,
  In *Kane, G.L. (ed.): Perspectives on supersymmetry II* 1-153
  [hep-ph/9709356].

\bibitem{Nilles:1983ge}
  H.~P.~Nilles,
  Phys.\ Rept.\  {\bf 110} (1984) 1.



\bibitem{Aad:2011ib}
  G.~Aad {\it et al.}  [ATLAS Collaboration],
  arXiv:1109.6572 [hep-ex].

\bibitem{ATLAS:2011ad}
  G.~Aad {\it et al.}  [ATLAS Collaboration],
  Phys.\ Rev.\ D {\bf 85} (2012) 012006
  [arXiv:1109.6606 [hep-ex]].

\bibitem{Aad:2011cwa}
  G.~Aad {\it et al.}  [ATLAS Collaboration],
  Phys.\  Lett.\ B {\bf 709} (2012) 137
  [arXiv:1110.6189 [hep-ex]].

\bibitem{Chatrchyan:2011ek}
  S.~Chatrchyan {\it et al.}  [CMS Collaboration],
  Phys.\ Rev.\ D {\bf 85} (2012) 012004
  [arXiv:1107.1279 [hep-ex]].


\bibitem{Chatrchyan:2011qs} 
  S.~Chatrchyan {\it et al.}  [CMS Collaboration],
  JHEP {\bf 1108}, 156 (2011)
  [arXiv:1107.1870 [hep-ex]].

\bibitem{Chatrchyan:2011zy}
  S.~Chatrchyan {\it et al.}  [CMS Collaboration],
  Phys.\ Rev.\ Lett.\  {\bf 107} (2011) 221804
  [arXiv:1109.2352 [hep-ex]].


\bibitem{moriond}
See the talk by Steve Lowette on behalf of the \texttt{Atlas} and 
\texttt{CMS} collaborations at the conference Rencontres de Moriond 2012
\url{http://indico.in2p3.fr/getFile.py/access?contribId=96&sessionId=5&resId=0&materialId=slides&confId=6001}.

\bibitem{Bechtle:2012zk}
  P.~Bechtle {\it et al.},
  arXiv:1204.4199 [hep-ph].

\bibitem{other}
  P.~Bechtle {\it et al.},
  Phys.\ Rev.\ D {\bf 84} (2011) 011701
  [arXiv:1102.4693 [hep-ph]];
O.~Buchmueller {\it et al.},
  Eur.\ Phys.\ J.\ C {\bf 72} (2012) 1878
  [arXiv:1110.3568 [hep-ph]];
B.~C.~Allanach,
  Phys.\ Rev.\ D {\bf 83} (2011) 095019
  [arXiv:1102.3149 [hep-ph]];
A.~Fowlie {\it et al.}, 
  Phys.\ Rev.\ D {\bf 85} (2012) 075012
  [arXiv:1111.6098 [hep-ph]].


\bibitem{Farrar:1978xj}
  G.~R.~Farrar and P.~Fayet,
  Phys.\ Lett.\ B {\bf 76} (1978) 575.

\bibitem{Haber:1997if}
  H.~E.~Haber,
  Nucl.\ Phys.\ Proc.\ Suppl.\  {\bf 62} (1998) 469
  [hep-ph/9709450].

\bibitem{Giudice:1998bp} 
  G.~F.~Giudice and R.~Rattazzi,
  Phys.\ Rept.\  {\bf 322}, 419 (1999)
  [hep-ph/9801271].


\bibitem{Paige:1999ui} 
  F.~E.~Paige and J.~D.~Wells,
  hep-ph/0001249.


\bibitem{Baer:2007eh}
  H.~Baer, E.~-K.~Park, X.~Tata and T.~T.~Wang,
  JHEP {\bf 0706} (2007) 033
  [hep-ph/0703024].

\bibitem{Conley:2011tq}
  J.~A.~Conley, H.~K.~Dreiner, L.~Glaser, M.~Kramer and J.~Tattersall,
  JHEP {\bf 1203} (2012) 042
  [arXiv:1110.1287 [hep-ph]].


\bibitem{Paige:2003mg}
  F.~E.~Paige, S.~D.~Protopopescu, H.~Baer and X.~Tata,
  hep-ph/0312045.

\bibitem{Allanach:2001kg}
  B.~C.~Allanach,
  Comput.\ Phys.\ Commun.\  {\bf 143} (2002) 305
  [hep-ph/0104145].

\bibitem{Djouadi:2002ze}
  A.~Djouadi, J.~-L.~Kneur and G.~Moultaka,
  Comput.\ Phys.\ Commun.\  {\bf 176} (2007) 426
  [hep-ph/0211331].

\bibitem{Porod:2003um}
  W.~Porod,
  Comput.\ Phys.\ Commun.\  {\bf 153} (2003) 275
  [hep-ph/0301101].


\bibitem{Konar:2010bi}
  P.~Konar, K.~T.~Matchev, M.~Park and G.~K.~Sarangi,
  Phys.\ Rev.\ Lett.\  {\bf 105} (2010) 221801
  [arXiv:1008.2483 [hep-ph]].

\bibitem{Berger:2008cq}
  C.~F.~Berger, J.~S.~Gainer, J.~L.~Hewett and T.~G.~Rizzo,
  JHEP {\bf 0902} (2009) 023
  [arXiv:0812.0980 [hep-ph]].

\bibitem{pMSSM}
For the pMSSM see Ref. \cite{Djouadi:2002ze} and
  P.~Bechtle, K.~Desch and P.~Wienemann,
  Comput.\ Phys.\ Commun.\  {\bf 174} (2006) 47
  [hep-ph/0412012].


\bibitem{Hall:1983id}
  L.~J.~Hall and M.~Suzuki,
  Nucl.\ Phys.\ B {\bf 231} (1984) 419.

\bibitem{Allanach:2003eb}
  B.~C.~Allanach, A.~Dedes and H.~K.~Dreiner,
  Phys.\ Rev.\ D {\bf 69} (2004) 115002
   [Erratum-ibid.\ D {\bf 72} (2005) 079902]
  [hep-ph/0309196].

\bibitem{Dreiner:1997uz}
  H.~K.~Dreiner,
  In *Kane, G.L. (ed.): Perspectives on supersymmetry II* 565-583
  [hep-ph/9707435].

\bibitem{Bhattacharyya:1997vv}
  G.~Bhattacharyya,
  In *Tegernsee 1997, Beyond the desert 1997* 194-201
  [hep-ph/9709395].

\bibitem{Barger:1989rk}
  V.~D.~Barger, G.~F.~Giudice and T.~Han,
  Phys.\ Rev.\  D {\bf 40} (1989) 2987.

\bibitem{Allanach:1999ic}
  B.~C.~Allanach, A.~Dedes and H.~K.~Dreiner,
  Phys.\ Rev.\  D {\bf 60} (1999) 075014
  [arXiv:hep-ph/9906209].

\bibitem{Hirsch:2000ef}
  M.~Hirsch {\it et al.},
  Phys.\ Rev.\ D {\bf 62} (2000) 113008
   [Erratum-ibid.\ D {\bf 65} (2002) 119901]
  [hep-ph/0004115].

\bibitem{Barbier:2004ez} 
  R.~Barbier {\it et al.},
  Phys.\ Rept.\  {\bf 420}, 1 (2005)
  [hep-ph/0406039].

\bibitem{Allanach:2008qq}
  B.~C.~Allanach {\it et al.},
  Comput.\ Phys.\ Commun.\  {\bf 180} (2009) 8
  [arXiv:0801.0045 [hep-ph]].

\bibitem{Allanach:1997sa}
  B.~C.~Allanach, H.~K.~Dreiner, P.~Morawitz and M.~D.~Williams,
  Phys.\ Lett.\ B {\bf 420} (1998) 307
  [hep-ph/9708495].

\bibitem{Bernhardt:2008mz}
  M.~A.~Bernhardt, H.~K.~Dreiner, S.~Grab and P.~Richardson,
  Phys.\ Rev.\ D {\bf 78} (2008) 015016
  [arXiv:0802.1482 [hep-ph]].


\bibitem{Dimopoulos:1988fr}
  S.~Dimopoulos, R.~Esmailzadeh, L.~J.~Hall and G.~D.~Starkman,
  Phys.\ Rev.\ D {\bf 41} (1990) 2099.

\bibitem{Dreiner:2000vf}
  H.~K.~Dreiner, P.~Richardson and M.~H.~Seymour,
  Phys.\ Rev.\ D {\bf 63} (2001) 055008
  [hep-ph/0007228].

\bibitem{Kao:2009fg}
  Y.~Kao and T.~Takeuchi,
  arXiv:0910.4980 [hep-ph].

\bibitem{Dreiner:2010ye}
  H.~K.~Dreiner, M.~Hanussek and S.~Grab,
  Phys.\ Rev.\ D {\bf 82} (2010) 055027
  [arXiv:1005.3309 [hep-ph]].

\bibitem{Dreiner:2012mx}
  H.~K.~Dreiner, K.~Nickel, F.~Staub and A.~Vicente,
  Phys.\  Rev.\  D 86, {\bf 015003} (2012)
  [arXiv:1204.5925 [hep-ph]].

\bibitem{Dreiner:2001kc}
  H.~K.~Dreiner, G.~Polesello and M.~Thormeier,
  Phys.\ Rev.\ D {\bf 65} (2002) 115006
  [hep-ph/0112228].

\bibitem{Dreiner:2006gu}
  H.~K.~Dreiner, M.~Kramer and B.~O'Leary,
  Phys.\ Rev.\ D {\bf 75} (2007) 114016
  [hep-ph/0612278].

\bibitem{Allanach:1999mh}
  B.~C.~Allanach, A.~Dedes and H.~K.~Dreiner,
  Phys.\ Rev.\ D {\bf 60} (1999) 056002
  [hep-ph/9902251].



\bibitem{Allanach:2006st}
  B.~C.~Allanach {\it et al.}, 
  Phys.\ Rev.\ D {\bf 75} (2007) 035002
  [hep-ph/0609263].

\bibitem{Preskill:1982cy}
  J.~Preskill, M.~B.~Wise and F.~Wilczek,
  Phys.\ Lett.\ B {\bf 120} (1983) 127.

\bibitem{Abbott:1982af}
  L.~F.~Abbott and P.~Sikivie,
  Phys.\ Lett.\ B {\bf 120} (1983) 133.

\bibitem{Dine:1982ah}
  M.~Dine and W.~Fischler,
  Phys.\ Lett.\ B {\bf 120} (1983) 137.

\bibitem{Covi:1999ty}
  L.~Covi, J.~E.~Kim and L.~Roszkowski,
  Phys.\ Rev.\ Lett.\  {\bf 82} (1999) 4180
  [hep-ph/9905212].

\bibitem{Ellis:2003dn}
  J.~R.~Ellis, K.~A.~Olive, Y.~Santoso and V.~C.~Spanos,
  Phys.\ Lett.\ B {\bf 588} (2004) 7
  [hep-ph/0312262].

\bibitem{Ellis:2006vu}
  J.~R.~Ellis, A.~R.~Raklev and O.~K.~Oye,
  JHEP {\bf 0610} (2006) 061
  [hep-ph/0607261].


\bibitem{Steffen:2008qp}
  F.~D.~Steffen,
  Eur.\ Phys.\ J.\ C {\bf 59} (2009) 557
  [arXiv:0811.3347 [hep-ph]].

\bibitem{Abazov:2008qu}
  V.~M.~Abazov {\it et al.}  [D0 Collaboration],
  Phys.\ Rev.\ Lett.\  {\bf 102} (2009) 161802
  [arXiv:0809.4472 [hep-ex]].


\bibitem{Aaltonen:2009kea}
  T.~Aaltonen {\it et al.}  [CDF Collaboration],
  Phys.\ Rev.\ Lett.\  {\bf 103} (2009) 021802
  [arXiv:0902.1266 [hep-ex]].



\bibitem{Raby:1997bpa} 
  S.~Raby,
  Phys.\ Lett.\ B {\bf 422}, 158 (1998)
  [hep-ph/9712254].


\bibitem{Baer:1998pg}
  H.~Baer, K.~-m.~Cheung and J.~F.~Gunion,
  Phys.\ Rev.\ D {\bf 59} (1999) 075002
  [hep-ph/9806361].


\bibitem{Raby:1998xr}
  S.~Raby and K.~Tobe,
  Nucl.\ Phys.\ B {\bf 539} (1999) 3
  [hep-ph/9807281].

\bibitem{deCampos:2007bn}
  F.~de Campos {\it et al.},
  JHEP {\bf 0805} (2008) 048
  [arXiv:0712.2156 [hep-ph]].

\bibitem{deCampos:2008av}
  F.~de Campos {\it et al.},
  Phys.\ Rev.\ D {\bf 77} (2008) 115025
  [arXiv:0803.4405 [hep-ph]].

\bibitem{DeCampos:2010yu}
  F.~De Campos {\it et al.},
  Phys.\ Rev.\ D {\bf 82} (2010) 075002
  [arXiv:1006.5075 [hep-ph]].

\bibitem{Desch:2010gi}
  K.~Desch, S.~Fleischmann, P.~Wienemann, H.~K.~Dreiner and S.~Grab,
  Phys.\ Rev.\ D {\bf 83} (2011) 015013
  [arXiv:1008.1580 [hep-ph]].


\bibitem{Dreiner:2011ft}
  H.~K.~Dreiner, M.~Hanussek, J.~-S.~Kim and C.~H.~Kom,
  Phys.\ Rev.\ D {\bf 84} (2011) 113005
  [arXiv:1106.4338 [hep-ph]].

\bibitem{kaplan2012}
  P.~W.~Graham, D.~E.~Kaplan, S.~Rajendran and P.~Saraswat,
  arXiv:1204.6038 [hep-ph].


\bibitem{Porod:2000hv}
  W.~Porod, M.~Hirsch, J.~Romao and J.~W.~F.~Valle,
  Phys.\ Rev.\ D {\bf 63} (2001) 115004
  [hep-ph/0011248].


\bibitem{Hirsch:2003fe}
  M.~Hirsch and W.~Porod,
  Phys.\ Rev.\ D {\bf 68} (2003) 115007
  [hep-ph/0307364].

\bibitem{Heister:2003hc}
  A.~Heister {\it et al.}  [ALEPH Collaboration],
  Eur.\ Phys.\ J.\ C {\bf 31} (2003) 327
  [hep-ex/0305071].


\bibitem{Arvanitaki:2005nq}
  A.~Arvanitaki {\it et al.},
  Phys.\ Rev.\ D {\bf 76} (2007) 055007
  [hep-ph/0506242].

\bibitem{Abazov:2007ht}
  V.~M.~Abazov {\it et al.}  [D0 Collaboration],
  Phys.\ Rev.\ Lett.\  {\bf 99} (2007) 131801
  [arXiv:0705.0306 [hep-ex]].

\bibitem{Aad:2011yf}
  G.~Aad {\it et al.}  [ATLAS Collaboration],
  Phys.\ Lett.\ B {\bf 701} (2011) 1
  [arXiv:1103.1984 [hep-ex]].

\bibitem{Masiero:1990uj}
  A.~Masiero and J.~W.~F.~Valle,
  Phys.\ Lett.\ B {\bf 251} (1990) 273.

\bibitem{Hirsch:2008ur}
  M.~Hirsch, A.~Vicente and W.~Porod,
  Phys.\ Rev.\ D {\bf 77} (2008) 075005
  [arXiv:0802.2896 [hep-ph]].

\bibitem{LopezFogliani:2005yw}
  D.~E.~Lopez-Fogliani and C.~Munoz,
  Phys.\ Rev.\ Lett.\  {\bf 97} (2006) 041801
  [hep-ph/0508297].

\bibitem{Bartl:2009an}
  A.~Bartl {\it et al.}, 
  JHEP {\bf 0905} (2009) 120
  [arXiv:0903.3596 [hep-ph]].



\bibitem{Dreiner:2009fi}
H.~K.~Dreiner and S.~Grab,
  Phys.\ Lett.\ B {\bf 679} (2009) 45
  [arXiv:0811.0200 [hep-ph]];
  H.~K.~Dreiner and S.~Grab,
  AIP Conf.\ Proc.\  {\bf 1200} (2010) 358
  [arXiv:0909.5407 [hep-ph]].

\bibitem{Butterworth:2008iy}
  J.~M.~Butterworth, A.~R.~Davison, M.~Rubin and G.~P.~Salam,
  Phys.\ Rev.\ Lett.\  {\bf 100} (2008) 242001
  [arXiv:0802.2470 [hep-ph]].

\bibitem{Butterworth:2009qa}
  J.~M.~Butterworth, J.~R.~Ellis, A.~R.~Raklev and G.~P.~Salam,
  Phys.\ Rev.\ Lett.\  {\bf 103} (2009) 241803
  [arXiv:0906.0728 [hep-ph]].



\bibitem{Allanach:2012vj}
  B.~C.~Allanach and B.~Gripaios,
  arXiv:1202.6616 [hep-ph].


\end{thebibliography}
\end{document}